\documentclass[12pt]{article}

\makeatletter

\def\paragraph{\@startsection{paragraph}{4}{\z@}{+2.00ex plus
 +1ex minus +.2ex}{1.5ex plus .2ex}{\it\normalsize}}

\def\section{\@startsection {section}{1}{\z@}{+3.0ex plus +1ex minus
  +.2ex}{2.3ex plus .2ex}{\normalsize\bf\boldmath}}
\def\subsection{\@startsection{subsection}{2}{\z@}{+2.5ex plus +1ex
minus +.2ex}{1.5ex plus .2ex}{\normalsize\bf\boldmath}}
\def\subsubsection{\@startsection{subsubsection}{3}{\z@}{+3.25ex plus
 +1ex minus +.2ex}{1.5ex plus .2ex}{\normalsize\it}}

\expandafter\ifx\csname mathrm\endcsname\relax\def\mathrm#1{{\rm #1}}\fi

\renewcommand{\theequation}{\thesection.\arabic{equation}}
\newcounter{saveeqn}

\@addtoreset{equation}{section}

\newcount\@tempcntc
\def\@citex[#1]#2{\if@filesw\immediate\write\@auxout{\string\citation{#2}}\fi
  \@tempcnta\z@\@tempcntb\m@ne\def\@citea{}\@cite{\@for\@citeb:=#2\do
    {\@ifundefined
       {b@\@citeb}{\@citeo\@tempcntb\m@ne\@citea
        \def\@citea{,\penalty\@m\ }{\bf ?}\@warning
       {Citation `\@citeb' on page \thepage \space undefined}}%
    {\setbox\z@\hbox{\global\@tempcntc0\csname
b@\@citeb\endcsname\relax}%
     \ifnum\@tempcntc=\z@ \@citeo\@tempcntb\m@ne
       \@citea\def\@citea{,\penalty\@m}
       \hbox{\csname b@\@citeb\endcsname}%
     \else
      \advance\@tempcntb\@ne
      \ifnum\@tempcntb=\@tempcntc
      \else\advance\@tempcntb\m@ne\@citeo
      \@tempcnta\@tempcntc\@tempcntb\@tempcntc\fi\fi}}\@citeo}{#1}}

\def\@citeo{\ifnum\@tempcnta>\@tempcntb\else\@citea
  \def\@citea{,\penalty\@m}%
  \ifnum\@tempcnta=\@tempcntb\the\@tempcnta\else
   {\advance\@tempcnta\@ne\ifnum\@tempcnta=\@tempcntb \else
\def\@citea{--}\fi
    \advance\@tempcnta\m@ne\the\@tempcnta\@citea\the\@tempcntb}\fi\fi}

\newcommand{\lsim}
{\mathrel{\raisebox{-.3em}{$\stackrel{\displaystyle <}{\sim}$}}}

\def\asymp#1%
{\mathrel{\raisebox{-.4em}{$\widetilde{\scriptstyle #1}$}}}

\def\Nequal#1%
{\mathrel{\raisebox{-.5em}{$\stackrel{=}{\scriptstyle\rm#1}$}}}
\newcommand{\dsl}[1]{\not \hspace{-0.7mm}#1}
\def\dsl{\mathpalette\make@slash}
\def\make@slash#1#2{\setbox\z@\hbox{$#1#2$}%
  \hbox to 0pt{\hss$#1/$\hss\kern-\wd0}\box0}

\def\beq{\begin{equation}}
\def\eeq{\end{equation}}
\def\beqar{\begin{eqnarray}}
\def\eeqar{\end{eqnarray}}
\def\barr#1{\begin{array}{#1}}
\def\earr{\end{array}}
\def\bfi{\begin{figure}}
\def\efi{\end{figure}}
\def\btab{\begin{table}}
\def\etab{\end{table}}
\def\bce{\begin{center}}
\def\ece{\end{center}}
\def\nn{\nonumber}

\def\text{\textstyle}

\def\al{\alpha}
\def\be{\beta}

\def\ga{\gamma}
\def\de{\delta}
\def\De{\Delta}
\def\eps{\epsilon}

\def\si{\sigma}

\def\refeq#1{\mbox{(\ref{#1})}}

\def\reffi#1{\mbox{Figure~\ref{#1}}}

\def\refta#1{\mbox{Table~\ref{#1}}}

\def\refse#1{\mbox{Section~\ref{#1}}}

\def\citere#1{\mbox{Ref.~\cite{#1}}}
\def\citeres#1{\mbox{Refs.~\cite{#1}}}

\newcommand{\GeV}{\unskip\,\mathrm{GeV}}
\newcommand{\MeV}{\unskip\,\mathrm{MeV}}

\newcommand{\rd}{{\mathrm{d}}}
\newcommand{\rE}{{\mathrm{E}}}

\newcommand{\Oa}{\mathswitch{{\cal{O}}(\alpha)}}

\newcommand{\M}{{\cal{M}}}

\def\mathswitchr#1{\relax\ifmmode{\mathrm{#1}}\else$\mathrm{#1}$\fi}

\newcommand{\PW}{\mathswitchr W}
\newcommand{\Pw}{\mathswitchr w}
\newcommand{\PZ}{\mathswitchr Z}

\newcommand{\PH}{\mathswitchr H}
\newcommand{\Pe}{\mathswitchr e}
\newcommand{\Pp}{\mathswitchr p}
\newcommand{\Pn}{\mathswitchr n}

\newcommand{\Pd}{\mathswitchr d}

\newcommand{\Pu}{\mathswitchr u}

\newcommand{\Ps}{\mathswitchr s}

\newcommand{\Pc}{\mathswitchr c}

\newcommand{\Pb}{\mathswitchr b}

\newcommand{\Pt}{\mathswitchr t}

\newcommand{\Pep}{\mathswitchr {e^+}}
\newcommand{\Pem}{\mathswitchr {e^-}}

\def\mathswitch#1{\relax\ifmmode#1\else$#1$\fi}

\newcommand{\MW}{\mathswitch {M_\PW}}

\newcommand{\MZ}{\mathswitch {M_\PZ}}
\newcommand{\MH}{\mathswitch {M_\PH}}
\newcommand{\Me}{\mathswitch {m_\Pe}}

\newcommand{\Md}{\mathswitch {m_\Pd}}
\newcommand{\Mu}{\mathswitch {m_\Pu}}
\newcommand{\Ms}{\mathswitch {m_\Ps}}
\newcommand{\Mc}{\mathswitch {m_\Pc}}
\newcommand{\Mb}{\mathswitch {m_\Pb}}
\newcommand{\Mt}{\mathswitch {m_\Pt}}

\newcommand{\sw}{\mathswitch {s_\Pw}}
\newcommand{\cw}{\mathswitch {c_\Pw}}

\newcommand{\GF}{\mathswitch {G_\mu}}

\def\solid{\raise.9mm\hbox{\protect\rule{1.1cm}{.2mm}}}
\def\dash{\raise.9mm\hbox{\protect\rule{2mm}{.2mm}}\hspace*{1mm}}

\def\sub{{\mathrm{sub}}}

\newcommand{\xbj}{x_{\mathrm{Bj}}}

\newcommand{\IPS}{\mathrm{IPS}}

\newcommand{\NC}{{\mathrm{NC}}}
\newcommand{\CC}{{\mathrm{CC}}}

\newcommand{\LL}{{\mathrm{LL}}}
\newcommand{\FSR}{{\mathrm{FSR}}}

\newcommand{\cut}{{\mathrm{cut}}}

\newcommand{\had}{{\mathrm{had}}}
\newcommand{\phot}{{\mathrm{phot}}}

\newcommand{\LAB}{{\mathrm{LAB}}}

\renewcommand{\max}{{\mathrm{max}}}

\newcommand{\MSbar}{\overline{\mathrm{MS}}}
\newcommand{\DIS}{{\mathrm{DIS}}}

\def\Li{\mathop{\mathrm{Li}_2}\nolimits}

\def\Re{\mathop{\mathrm{Re}}\nolimits}

\def\draftdate{\relax}
\def\mda{\relax}
\def\mua{\relax}
\def\mla{\relax}
\def\draft{
\def\thtystars{******************************}
\def\sixtystars{\thtystars\thtystars}
\typeout{}
\typeout{\sixtystars**}
\typeout{* Draft mode!
         For final version remove \protect\draft\space in source file *}
\typeout{\sixtystars**}
\typeout{}
\def\draftdate{\today}
\def\mua{\marginpar[\boldmath\hfil$\uparrow$]%
                   {\boldmath$\uparrow$\hfil}%
                    \typeout{marginpar: $\uparrow$}\ignorespaces}
\def\mda{\marginpar[\boldmath\hfil$\downarrow$]%
                   {\boldmath$\downarrow$\hfil}%
                    \typeout{marginpar: $\downarrow$}\ignorespaces}
\def\mla{\marginpar[\boldmath\hfil$\rightarrow$]%
                   {\boldmath$\leftarrow $\hfil}%
                    \typeout{marginpar: $\leftrightarrow$}\ignorespaces}
\def\Mua{\marginpar[\boldmath\hfil$\Uparrow$]%
                   {\boldmath$\Uparrow$\hfil}%
                    \typeout{marginpar: $\uparrow$}\ignorespaces}
\def\Mda{\marginpar[\boldmath\hfil$\Downarrow$]%
                   {\boldmath$\Downarrow$\hfil}%
                    \typeout{marginpar: $\downarrow$}\ignorespaces}
\def\Mla{\marginpar[\boldmath\hfil$\Rightarrow$]%
                   {\boldmath$\Leftarrow $\hfil}%
                    \typeout{marginpar: $\leftrightarrow$}\ignorespaces}
\overfullrule 5pt
\oddsidemargin -15mm
\marginparwidth 29mm
}

\def\stars{\strut\leaders\hbox{*}\hfill\strut}
\def\starline{\hfil\strut\hfil\hbox to \textwidth {\stars}\hfil}

\begin{document}
\thispagestyle{empty}
\def\thefootnote{\fnsymbol{footnote}}
\setcounter{footnote}{1}
\null
\draftdate\hfill MPP-2005-28 \\
\strut\hfill hep-ph/0509084
\vfill
\begin{center}
{\Large \bf\boldmath{
Electroweak higher-order effects and theoretical uncertainties
in deep-inelastic neutrino scattering}
\par} 
\vspace{1.2cm}
{\large
{\sc K.-P.~O.~Diener$^1$, S.~Dittmaier$^2$ and W.~Hollik$^2$} } \\[1cm]
$^1$ {\it Paul-Scherrer-Institut, W\"urenlingen und Villigen\\
CH-5232 Villigen PSI, Switzerland} \\[0.5cm]
$^2$ {\it Max-Planck-Institut f\"ur Physik 
(Werner-Heisenberg-Institut) \\
D-80805 M\"unchen, Germany}
\par \vskip 1em
\end{center}\par
\vskip 1.5cm {\bf Abstract:} \par 
A previous calculation of electroweak ${\cal O}(\alpha)$ corrections
to deep-inelastic neutrino scattering, as e.g.\ measured by NuTeV and 
NOMAD, is supplemented by higher-order effects. 
In detail, we take into account universal two-loop effects from 
$\De\alpha$ and $\De\rho$ as well as higher-order final-state photon
radiation off muons in the structure function approach.
Moreover, we make use of the recently released ${\cal O}(\alpha)$-improved
parton distributions MRST2004QED and identify the relevant QED
factorization scheme, which is DIS like.
As a technical byproduct, we describe slicing and subtraction techniques
for an efficient calculation of a new type of real corrections that are
induced by the generated photon distribution.
A numerical discussion of the higher-order effects suggests that the
remaining theoretical uncertainty from unknown electroweak 
corrections is dominated by non-universal two-loop effects
and is of the order 0.0003 when translated into a shift in
$\sin^2\theta_\PW=1-\MW^2/\MZ^2$. The ${\cal O}(\alpha)$ corrections
implicitly included in the parton distributions lead to a shift of
about $0.0004$.
\par
\vskip 3cm
\noindent
September 2005    
\null
\setcounter{page}{0}
\clearpage
\def\thefootnote{\arabic{footnote}}
\setcounter{footnote}{0}

\clearpage

\section{Introduction}
\label{sec:intro}

\begin{sloppypar}
In 2002 the NuTeV collaboration has deduced the on-shell
weak mixing angle, $\sin^2\theta_\PW=1-\MW^2/\MZ^2$, from the
ratio of neutral- to charged-current cross-sections 
measured in deep-inelastic neutrino scattering \cite{Zeller:2001hh}.
The NuTeV result on $\sin^2\theta_\PW$ is about $3\si$ at variance with
the overall fit \cite{unknown:2004qh} of the Standard Model (SM) to
precision observables, a fact that is known as ``the NuTeV anomaly''.
In spite of many attempts to understand or to resolve this anomaly 
in or beyond 
\cite{Davidson:2001ji,Bernstein:2002sa,Gambino:2002xp,Giunti:2002nh,%
Londergan:2004nr}
the SM, the issue is still not settled.
Inside the SM, uncertainties originating from parton distributions
\cite{Bernstein:2002sa,Gambino:2002xp,Londergan:2004nr,Zeller:2002du,%
McFarland:2003jw,Martin:2004dh}
(isospin violation, strange-sea asymmetry), 
from nuclear effects 
\cite{Bernstein:2002sa,Londergan:2004nr,McFarland:2003jw,Miller:2002xh}, 
and from QCD 
\cite{Davidson:2001ji,McFarland:2003jw,Kretzer:2003iu}
and 
electroweak \cite{Marciano:pb,Bardin:1986bc,Diener:2003ss,Arbuzov:2004zr}
radiative corrections
significantly influence the NuTeV result.
In view of the recent progress on these subjects, a careful reanalysis
of the NuTeV result, including its error assessment, would be welcome.
In this context it is very interesting that the NOMAD
collaboration plans to extract $\sin^2\theta_\PW$ from
their data on neutrino scattering \cite{Petti:2004wy}.
\end{sloppypar}

In our previous work \cite{Diener:2003ss}, we performed a calculation
of the electroweak corrections of ${\cal O}(\alpha)$ to neutral- and
charged-current deep-inelastic neutrino scattering and compared our
results with the older calculation \cite{Bardin:1986bc} which was
used in the NuTeV data analysis. Unfortunately, we could not fully 
reproduce the results of \citere{Bardin:1986bc}, which was in part
certainly due to the incompletely known setup and input. However, a
comparison with a more recent recalculation \cite{Arbuzov:2004zr},
which is in line with \citere{Bardin:1986bc}, suggests that the
still remaining differences are due to different kinematical
approximations made in \citeres{Bardin:1986bc,Arbuzov:2004zr}
compared to our calculation. While our calculation \cite{Diener:2003ss}
is a pure ${\cal O}(\alpha)$ calculation without further approximations,
in \citeres{Bardin:1986bc,Arbuzov:2004zr} the momentum transfer of
the $\PW$ and $\PZ$ bosons was neglected, i.e.\ the approximation of
a four-fermion contact interaction was made.
Moreover, we concluded in \citere{Diener:2003ss} that the estimate
of remaining theoretical uncertainties from unknown electroweak
higher orders was too optimistic.

In this paper, we refine our previous calculation \cite{Diener:2003ss} 
of the electroweak corrections to deep-inelastic neutrino scattering
by including leading effects beyond ${\cal O}(\alpha)$, in order to
further reduce the theoretical uncertainty in the electroweak sector.
In detail, after a short summary of our conventions in \refse{se:kinconv}
we work out the leading universal corrections originating
from the running of the electromagnetic coupling and from the $\rho$
parameter in various input-parameter schemes in \refse{se:llho}, 
in order to reduce the scheme dependence. Moreover, in \refse{se:llfsr}
we include the leading logarithmic
effects beyond ${\cal O}(\alpha)$ that result from collinear
final-state radiation (FSR) in the structure-function approach.
Finally, we make use of the recently released ${\cal O}(\alpha)$-improved
parton distribution functions (PDFs) of the MRST collaboration
\cite{Martin:2004dh}, which requires two kind of refinements,
as explained in \refse{se:OaPDFs}.
Firstly, the factorization scheme for the ${\cal O}(\alpha)$
corrections has to be chosen in a way that is consistent with the
treatment of ${\cal O}(\alpha)$ corrections used in the fit of the PDFs.
Secondly, the ${\cal O}(\alpha)$ driven evolution of the PDFs leads
to an induced photon distribution, so that a new type of real
corrections with a photon in the initial state has to be considered.
For a proper technical treatment of the photon-induced processes we have 
worked out different variants of phase-space slicing and dipole
subtraction, as described in the Appendix in detail.
Section~\ref{se:results} contains our numerical results, which
comprise a discussion of the various higher-order corrections
both to integrated and differential cross sections. Moreover,
we derive an estimate of the remaining theoretical uncertainties
from still missing electroweak effects there.
Our conclusions are given in \refse{se:concl}.


\section{Kinematics and conventions}
\label{se:kinconv}

In this section we briefly repeat our conventions of \citere{Diener:2003ss}
as far as relevant in the sequel.
We consider the neutral-current (NC) and charged-current (CC) parton processes
\beqar
\mbox{NC:} \quad \nu_\mu(p_l) + q(p_q) &\;\to\;& \nu_\mu(k_l) + q(k_q), 
\qquad q=\Pu,\Pd,\Ps,\Pc,\bar\Pu,\bar\Pd,\bar\Ps,\bar\Pc,
\label{eq:ncprocesses}
\\
\mbox{CC:} \quad \nu_\mu(p_l) + q(p_q) &\;\to\;& \mu^-(k_l) + q'(k_q),
\qquad q=\Pd,\Ps,\bar\Pu,\bar\Pc,
\quad q'=\Pu,\Pc,\bar\Pd,\bar\Ps,
\label{eq:ccprocesses}
\eeqar
and the processes with all particles replaced by their antiparticles.
The assignment of momenta is indicated in parentheses.
At lowest order the former processes proceed via Z-boson exchange, the latter
via W-boson exchange.
CP symmetry implies that
the (parton) cross sections for $\nu_\mu$--quark and $\nu_\mu$--antiquark
scattering are equal to the ones of 
$\bar\nu_\mu$--antiquark and $\bar\nu_\mu$--quark scattering,
respectively.

In the discussion of numerical results, we concentrate on
the NC to CC cross-section ratio~\cite{LlewellynSmith:ie}
\beq
R^\nu = \frac{\si_\NC^\nu(\nu_\mu N\to\nu_\mu X)}
{\si_\CC^\nu(\nu_\mu N\to\mu^- X)},
\label{eq:Rnu}
\eeq
but our calculation is applicable to any combination of NC and CC cross
section involving incoming neutrinos or anti-neutrinos, such as the
ratio
\beq
R^- = \frac{\si_\NC^\nu(\nu_\mu N\to\nu_\mu X)
-\si_\NC^{\bar\nu}(\bar\nu_\mu N\to\bar\nu_\mu X)}
{\si_\CC^\nu(\nu_\mu N\to\mu^- X)-\si_\CC^{\bar\nu}(\bar\nu_\mu N\to\mu^+ X)},
\eeq
as proposed by Paschos and Wolfenstein \cite{Paschos:1972kj}.

Throughout the calculation, we neglect the muon mass and the parton masses
whenever possible. The quark masses are only kept as regulators
for mass singularities, which appear as mass logarithms in the
(photonic) corrections; the quark-mass logarithms are absorbed into
the PDFs as described in \refse{se:PDF}.
The muon mass is not only required as regulator for collinear
final-state radiation but also for a proper description of 
forward scattering in the (loop-induced) $\gamma\nu_\mu\bar\nu_\mu$ vertex,
as described in \citere{Diener:2003ss} in detail.

The momentum of the incoming (anti-)quark $q$ is related to the
total nucleon momentum $P_N$ by the usual scaling relation
\beq
p_q^\mu = x P_N^\mu,
\eeq
which holds in the centre-of-mass (CM) 
frame of the partonic system where the nucleon
mass $M_N$ is negligible. The variable $x$ is the usual momentum
fraction, restricted by $0<x<1$.
Since deep-inelastic neutrino scattering usually is accessed 
in fixed-target
experiments (as NuTeV), we take the energy $E_\nu^{\LAB}$ of the
incoming (anti-)neutrino beam to define the incoming momentum $p_l$
in the rest frame of the nucleon, called LAB in the following. 
Thus, the squared partonic CM energy is given by
\beq
s = (p_l+p_q)^2 = 2x M_N E_\nu^{\LAB},
\label{eq:s_EnuLAB}
\eeq
up to terms of higher order in the nucleon mass.
Moreover, we define the Bjorken variable 
\beq
\xbj = \frac{-(p_l-k_l)^2}{2P_N (p_l-k_l)} 
= \frac{-(p_l-k_l)^2}{2M_N E_{\had+\phot}^{\LAB}}
\eeq
and the energy ratio $y$ via
\beq
y= \frac{E_{\had+\phot}^{\LAB}}{E_\nu^{\LAB}},
\eeq
where $E_{\had+\phot}^{\LAB}$ is the energy (in the LAB frame)
deposited by hadrons and photons in the detector.
In lowest order $\xbj$ and $x$ coincide, and
$E_{\had+\phot}^{\LAB}$ is identical with the energy of
the outgoing quark; in the presence of photon (or gluon) radiation
this is not the case anymore.


\section{\boldmath{Universal electroweak corrections beyond 
${\cal O}(\alpha)$}}
\label{se:llho}

The integrated CC and NC cross sections at parton level,
including ${\cal O}(\alpha)$ corrections, can be written as
\beqar
\sigma_{\CC,1}\Big|^\IPS &=& 
\sigma_{\CC,0}\Big|^\IPS \left(1+\delta_{\CC,1}\Big|^\IPS\right),
\nn\\
\sigma_{\NC,1}^\tau\Big|^\IPS &=& 
\sigma_{\NC,0}^\tau\Big|^\IPS 
\left(1+\delta_{\NC,1}^\tau\Big|^\IPS\right),
\eeqar
where $\sigma_{\CC,0}$ and $\sigma_{\NC,0}^\tau$ denote the partonic 
lowest-order cross sections, and $\tau=\pm$ refers to the chirality
of the quark line. The input-parameter scheme (IPS) is
indicated by an upper index which takes the value
$\alpha(0)$, $\alpha(\MZ)$, or $\GF$ for the
respective scheme. The precise definition of these IPS is
given in Section~4.2 of \citere{Diener:2003ss}.
Roughly speaking, the whole SM input consists of all particle
masses together with the chosen $\alpha$ for the electromagnetic 
coupling.
The $\alpha(0)$-scheme corresponds to the ``complete on-shell
renormalization scheme'' as described, e.g., in 
\citeres{Denner:1993kt,Bohm:rj}.
The relative ${\cal O}(\alpha)$ corrections $\delta_{\CC,1}$ and
$\delta_{\NC,1}^\tau$ are proportional to the value of $\alpha$ 
in the corresponding IPS, where $\alpha_{\GF}=\sqrt{2}\GF\MW^2\sw^2/\pi$
in the $\GF$-scheme.
Apart from this factor the relative corrections differ by
constant terms,
\beqar
\delta_1\Big|^{\alpha(\MZ)} &=& \frac{\alpha(\MZ)}{\alpha(0)}
\left[\delta_1\Big|^{\alpha(0)}-2\Delta\alpha(\MZ)\right],
\nn\\
\delta_1\Big|^{\GF} &=& \frac{\alpha_{\GF}}{\alpha(0)}
\left[\delta_1\Big|^{\alpha(0)}-2\Delta r_1\right].
\eeqar
Here
\beq
\Delta\alpha(\MZ) = 
\left.\left[\Pi^{\gamma\gamma}(0)-\Re\{\Pi(\MZ^2)\}\right]
\right|_{f\ne\mathrm{top}}
\approx
\frac{\alpha(0)}{3\pi}\sum_{f\ne\mathrm{top}} N^{\mathrm{c}}_f Q_f^2
\left[\ln\biggl(\frac{\MZ^2}{m_f^2}\biggr)-\frac{5}{3}\right]
\eeq
describes the running of the electromagnetic coupling from the scale
$Q=0$ to $Q=\MZ$, induced by the light fermions $f$ (all but the top 
quark) with colour factor $N^{\mathrm{c}}_f$, electric charge $Q_f$, and
mass $m_f$.
The quantity $\Pi^{\gamma\gamma}(Q^2)$ is the photonic vacuum
polarization in the convention of \citere{Denner:1993kt}.
The quantity $\Delta r_1$ is the pure one-loop correction to
muon decay \cite{Sirlin:1980nh}, as defined by
\beq
\MW^2 \left(1-\frac{\MW^2}{\MZ^2}\right) =
\frac{\pi\alpha(0)}{\sqrt{2}\GF}
\left[1+\Delta r_1(\alpha(0),\MW,\MZ,\MH,m_f)\right].
\eeq
Note that $\Delta r_1$ implicitly contains large contributions
from $\Delta\alpha(\MZ)\sim6\%$ and the (one-loop) correction 
$(\cw^2/\sw^2)\Delta\rho_1\sim3\%$
induced by the $\rho$-parameter,
\beq
\Delta r_1 = \Delta\alpha(\MZ) 
- \frac{\cw^2}{\sw^2}\Delta\rho_1\Big|^{\alpha(0)} 
+ \Delta r_{\mathrm{rem}},
\eeq
where
\beq
\Delta\rho_1\Big|^\alpha = \frac{3\alpha}{16\pi\sw^2}\frac{\Mt^2}{\MW^2}.
\eeq
The remainder $\Delta r_{\mathrm{rem}}$ comprises the non-universal
one-loop corrections to muon decay and is of the size of $\sim 1\%$.

The leading universal two-loop effects induced by 
$\Delta\alpha(\MZ)$ and by corrections $\Delta\rho$ in the parameter
$\rho=1/(1-\Delta\rho)$ can be included as described in
\citere{Consoli:1989pc}. The procedure is as follows.
Starting from lowest-order cross sections in the $\alpha(0)$-scheme,
all orders $\Delta\alpha(\MZ)^n$ can be effectively included upon
substituting $\alpha(0)\to\alpha(\MZ)=\alpha(0)/[1-\Delta\alpha(\MZ)]$
as electromagnetic coupling.
The effects from $\Delta\rho$ can be absorbed by replacing $\sw^2$
and $\cw^2$ by appropriate modifications
\beq
\bar\sw^2 = \sw^2+\cw^2\Delta\rho, \qquad \bar\cw^2=1-\bar\sw^2.
\eeq
This recipe is correct up to ${\cal O}(\Delta\rho^2)$.
As shown in \citere{Consoli:1989pc},
the introduction of $\alpha(\MZ)$ and $\bar\sw^2$,
$\bar\cw^2$ also correctly
reproduces the correct terms of ${\cal O}(\Delta\alpha(\MZ)\Delta\rho)$
in processes with four light external fermions.
Note that in ${\cal O}(\Delta\rho^2)$ both one- and two-loop corrections
to $\Delta\rho$ become relevant; explicitly we use the result
\beq
\Delta\rho = 3x_\Pt\left[1+\rho^{(2)}\left(\MH^2/\Mt^2\right)x_\Pt\right]
\biggl[1-\frac{2\alpha_{\mathrm{s}}}{9\pi}(\pi^2+3) \biggr],
\qquad
3x_\Pt = \frac{3\sqrt{2}\GF\Mt^2}{16\pi^2} = 
\Delta\rho_1\Big|^{\GF},
\eeq
with the function $\rho^{(2)}$ given in Eq.~(12) of
\citere{Fleischer:1993ub}.
In order to avoid double-counting, the one-loop effects of
$\Delta\alpha(\MZ)$ and $\Delta\rho$ have to be subtracted from
the full one-loop corrections $\delta_1\Big|^{\alpha(0)}$
introduced above.

The above recipe can be easily modified to the other two IPS.
In the $\alpha(\MZ)$-scheme, the replacement $\alpha(0)\to\alpha(\MZ)$
has already been done, so that the procedure reduces to the
substitutions $\sw\to\bar\sw$ and $\cw\to\bar\cw$.
In the $\GF$-scheme, one has to recall that
$\alpha_{\GF}$ effectively involves a factor 
$\alpha(\MZ)\sw^2$, so that the recipe reads
$\alpha_{\GF}\to\alpha_{\GF}\bar\sw^2/\sw^2$,
$\sw\to\bar\sw$, and $\cw\to\bar\cw$; as a net effect
the combination $\alpha_{\GF}/\sw^2$ is not modified.

The lowest-order CC cross section is proportional to
$\alpha^2/\sw^4$. Thus, using the above recipes we find for
the corrected cross sections, which include the full ${\cal O}(\alpha)$
corrections as well as two-loop improvements from 
$\Delta\alpha\equiv\Delta\alpha(\MZ)$ and $\Delta\rho$, the results
\beqar
\label{eq:cc_llho_alp0}
\sigma_{\CC,1+\LL^2}\Big|^{\alpha(0)} &=& 
\sigma_{\CC,0}\Big|^{\alpha(0)} 
\Biggl[ \frac{\sw^4}{\bar\sw^4}\frac{1}{(1-\Delta\alpha)^2}
+\delta_{\CC,1}\Big|^{\alpha(0)}
-2\Delta\alpha+2\frac{\cw^2}{\sw^2}\Delta\rho_1\Big|^{\alpha(0)} \Biggr]
\nn\\
&=&
\sigma_{\CC,0}\Big|^{\alpha(0)} 
\Biggl[ 1 +\delta_{\CC,1}\Big|^{\alpha(0)}
+2\frac{\cw^2}{\sw^2}\left(\Delta\rho_1\Big|^{\alpha(0)}-\Delta\rho\right)
\nn\\
&& \hspace*{4.5em} {}
+3\Delta\alpha^2-4\frac{\cw^2}{\sw^2}\Delta\alpha\Delta\rho
+3\frac{\cw^4}{\sw^4}\Delta\rho^2
\Biggr] \;+\; \dots,
\\[1em]
\label{eq:cc_llho_alpmz}
\sigma_{\CC,1+\LL^2}\Big|^{\alpha(\MZ)} &=& 
\sigma_{\CC,0}\Big|^{\alpha(\MZ)} 
\Biggl[ \frac{\sw^4}{\bar\sw^4}
+\delta_{\CC,1}\Big|^{\alpha(\MZ)}
+2\frac{\cw^2}{\sw^2}\Delta\rho_1\Big|^{\alpha(\MZ)} \Biggr]
\nn\\
&=&
\sigma_{\CC,0}\Big|^{\alpha(\MZ)} 
\Biggl[ 1 +\delta_{\CC,1}\Big|^{\alpha(\MZ)}
+2\frac{\cw^2}{\sw^2}
  \left(\Delta\rho_1\Big|^{\alpha(\MZ)}-\Delta\rho\right)
+3\frac{\cw^4}{\sw^4}\Delta\rho^2
\Biggr] 
\nn\\
&& {} \;+\; \dots,
\\[1em]
\label{eq:cc_llho_gf}
\sigma_{\CC,1+\LL^2}\Big|^{\GF} &=& 
\sigma_{\CC,0}\Big|^{\GF} 
\left[1+\delta_{\CC,1}\Big|^{\GF}\right]
=\sigma_{\CC,1}\Big|^{\GF}.
\label{eq:CCllho_GF}
\eeqar
For the $\alpha(0)$- and $\alpha(\MZ)$-schemes the two given
expressions are equal up to non-leading two-loop and
leading three-loop terms. 
Equation~\refeq{eq:CCllho_GF} reflects the well-known fact
that the $\GF$-scheme, which employs $\GF$, $\MW$, and $\MZ$
as independent parameters, effectively absorbs the leading
effects from $\Delta\alpha(\MZ)$ and $\Delta\rho$ in the
coupling of W~bosons to fermions.

The lowest-order NC cross section is proportional to
$\alpha^2(g_{qqZ}^\tau)^2/(\cw\sw)^2$, where the coupling of the Z~boson
to a quark $q$ with chirality $\tau=\pm$ is given by
\beq
g^\tau_{qqZ} = -\frac{\sw}{\cw}Q_q+\frac{I_q^3}{\cw\sw}\,\de_{\tau-},
\eeq
with $I_q^3=\pm\frac{1}{2}$ denoting the third component of the weak
isospin of a left-handed quark.
In order to absorb corrections from $\Delta\rho$, we define
the modified coupling factors
\beq
\bar g^\tau_{qqZ} = g^\tau_{qqZ}\Big|_{\sw\to\bar\sw,\cw\to\bar\cw}.
\eeq
The ${\cal O}(\alpha)$-corrected NC cross sections, including the
leading two-loop improvements, thus, read
\beqar
\label{eq:nc_llho_alp0}
\sigma^\tau_{\NC,1+\LL^2}\Big|^{\alpha(0)} &=& 
\sigma^\tau_{\NC,0}\Big|^{\alpha(0)} \Biggl[ 
\biggl(\frac{\bar g^\tau_{qqZ}\cw\sw}{g^\tau_{qqZ}\bar\cw\bar\sw}\biggr)^2
\frac{1}{(1-\Delta\alpha)^2}
+\delta^\tau_{\NC,1}\Big|^{\alpha(0)}
-2\Delta\alpha
\nn\\
&& \hspace*{5em} {}
-2\Delta\rho_1\Big|^{\alpha(0)} 
  \Biggl(1-\frac{\cw I_q^3}{g^\tau_{qqZ}\sw^3}\,\de_{\tau-} \Biggr)
\Biggr]
\nn\\
&=&
\sigma^\tau_{\NC,0}\Big|^{\alpha(0)} 
\Biggl[ 1 +\delta_{\NC,1}\Big|^{\alpha(0)}
-2\left(\Delta\rho_1\Big|^{\alpha(0)}-\Delta\rho\right) 
  \Biggl(1-\frac{\cw I_q^3}{g^\tau_{qqZ}\sw^3}\,\de_{\tau-} \Biggr)
\nn\\
&& \hspace*{5em} {}
+3\Delta\alpha^2
+4\Delta\alpha\Delta\rho
  \Biggl(1-\frac{\cw I_q^3}{g^\tau_{qqZ}\sw^3}\,\de_{\tau-} \Biggr)
\nn\\
&& \hspace*{5em} {}
+\Delta\rho^2\left(3+\frac{\cw^2}{4(g^\tau_{qqZ})^2\sw^6}\,\de_{\tau-}
+\frac{2\cw(1-3\sw^2)I_q^3}{g^\tau_{qqZ}\sw^5}\,\de_{\tau-} \right)
\Biggr]
\nn\\
&& {} \;+\; \dots,
\\[1em]
\label{eq:nc_llho_alpmz}
\sigma^\tau_{\NC,1+\LL^2}\Big|^{\alpha(\MZ)} &=& 
\sigma^\tau_{\NC,0}\Big|^{\alpha(\MZ)}  \Biggl[ 
\biggl(\frac{\bar g^\tau_{qqZ}\cw\sw}{g^\tau_{qqZ}\bar\cw\bar\sw}\biggr)^2
+\delta^\tau_{\NC,1}\Big|^{\alpha(\MZ)}
\nn\\
&& \hspace*{5em} {}
-2\Delta\rho_1\Big|^{\alpha(\MZ)} 
  \Biggl(1-\frac{\cw I_q^3}{g^\tau_{qqZ}\sw^3}\,\de_{\tau-} \Biggr)
\Biggr]
\nn\\
&=&
\sigma^\tau_{\NC,0}\Big|^{\alpha(\MZ)} 
\Biggl[ 1 +\delta_{\NC,1}\Big|^{\alpha(\MZ)}
-2\left(\Delta\rho_1\Big|^{\alpha(\MZ)}-\Delta\rho\right) 
  \Biggl(1-\frac{\cw I_q^3}{g^\tau_{qqZ}\sw^3}\,\de_{\tau-} \Biggr)
\nn\\
&& \hspace*{5em} {}
+\Delta\rho^2\left(3+\frac{\cw^2}{4(g^\tau_{qqZ})^2\sw^6}\,\de_{\tau-}
+\frac{2\cw(1-3\sw^2)I_q^3}{g^\tau_{qqZ}\sw^5}\,\de_{\tau-} \right)
\Biggr]
\nn\\
&& {} \;+\; \dots,
\\[1em]
\label{eq:nc_llho_gf}
\sigma^\tau_{\NC,1+\LL^2}\Big|^{\GF} &=& 
\sigma^\tau_{\NC,0}\Big|^{\GF} \Biggl[
\biggl(\frac{\bar g^\tau_{qqZ}\cw\bar\sw}{g^\tau_{qqZ}\bar\cw\sw}\biggr)^2
+\delta^\tau_{\NC,1}\Big|^{\GF}
-\frac{2}{\sw^2}\Delta\rho_1\Big|^{\GF}
\left(1-\frac{\cw I_q^3}{g^\tau_{qqZ}\sw}\,\de_{\tau-}\right)
\Biggr]
\label{eq:NCllho_GF}
\nn\\
&=& \sigma^\tau_{\NC,0}\Big|^{\GF} \Biggl[
1+\delta^\tau_{\NC,1}\Big|^{\GF}
-\frac{2}{\sw^2}\left(\Delta\rho_1\Big|^{\GF}-\Delta\rho\right)
\left(1-\frac{\cw I_q^3}{g^\tau_{qqZ}\sw}\,\de_{\tau-}\right)
\Biggr]
\nn\\
&& \hspace*{5em} {}
+\frac{\Delta\rho^2}{\sw^4} 
\left(1+2\sw^2+\frac{\cw^2}{4(g^\tau_{qqZ})^2\sw^2}\,\de_{\tau-}
-\frac{2\cw(1+\sw^2)I_q^3}{g^\tau_{qqZ}\sw}\,\de_{\tau-}\right)
\nn\\
&& {} \;+\; \dots.
\eeqar

\section{Higher-order final-state radiation off muons}
\label{se:llfsr}

In the CC reaction, where a muon appears in the final state,
the emission of photons collinear to the outgoing muon
leads to corrections that are enhanced by large logarithms
of the form $\alpha\ln(m_\mu^2/Q^2)$ with $Q$ denoting a
scale characteristic for the process.
The KLN theorem \cite{KLN} guarantees that these logarithms
cancel if photons collinear to the muon are treated fully inclusively.
Since we, however, apply a phase-space cut on the energy
of the outgoing muon (or equivalently on the remaining total 
hadronic+photonic energy), 
contributions enhanced by these logarithms survive.
The first-order logarithm $\alpha\ln(m_\mu^2/Q^2)$ is, of course,
implicitly contained in the full ${\cal O}(\alpha)$ correction,
so that $Q$ is unambiguously fixed in this order. 
However, it is desirable to control the logarithmically enhanced
corrections beyond ${\cal O}(\alpha)$. This can be done in the
so-called structure function approach \cite{sf}, where
these logarithms are derived from the universal factorization
of the related mass singularity.
The incorporation of the mass-singular logarithms takes the
form of a convolution integral over the lowest-order cross section
$\sigma_{\CC,0}$,
\beq
  \sigma_{\CC,\LL\FSR} =
  \int \rd\sigma_{\CC,0}(p_l,p_q;k_l,k_q)
  \int^1_0 \rd z \, \Gamma_{\mu\mu}^{\LL}(z,Q^2) \,
  \Theta_{\cut}(z k_l),
\label{eq:llfsr}
\eeq
where the step function $\Theta_{\cut}$ is equal to 1 if the event
passes the cut on the rescaled muon momentum $z k_l$ and 0 otherwise.
The variable $z$ is the momentum fraction describing the muon energy
loss by collinear photon emission.
This treatment, thus, corresponds to the experimental situation in
a fixed-target experiment where photons in the final state are not
distinguished from the hadrons that shower in the detector, but
the muon is analyzed as an isolated particle.
For the structure function $\Gamma_{\mu\mu}^{\LL}(z,Q^2)$
we take into account terms up to ${\cal O}(\al^3)$ improved by
the well-known exponentiation of the soft-photonic parts \cite{sf},
\newcommand{\betamu}{\be_\mu}
\newcommand{\betae}{\be_\Pe}
\beqar
  \Gamma_{\mu\mu}^{\LL}(z,Q^2) &=&    
    \frac{\exp\left(-\frac{1}{2}\betamu\gamma_{\rE} +
        \frac{3}{8}\betamu\right)}
{\Gamma\left(1+\frac{1}{2}\betamu\right)}
    \frac{\betamu}{2} (1-z)^{\frac{\betamu}{2}-1} - \frac{\betamu}{4}(1+z) 
\nn\\
&&  {} - \frac{\betamu^2}{32} \biggl\{ \frac{1+3z^2}{1-z}\ln(z)
    + 4(1+z)\ln(1-z) + 5 + z \biggr\}
\nn\\
&&  {} - \frac{\betamu^3}{384}\biggl\{
      (1+z)\left[6\Li(z)+12\ln^2(1-z)-3\pi^2\right] 
\nn\\
&& \quad\quad {}
+\frac{1}{1-z}\biggl[ \frac{3}{2}(1+8z+3z^2)\ln(z) 
+6(z+5)(1-z)\ln(1-z)
\nn\\
&& \quad\quad\quad {}
+12(1+z^2)\ln(z)\ln(1-z)-\frac{1}{2}(1+7z^2)\ln^2(z)
\nn\\
&& \quad\quad\quad  {}
+\frac{1}{4}(39-24z-15z^2)\biggr] \biggr\}
\nn\\
&& {} 
+\frac{\betamu\betae}{48}\left(\frac{1+z^2}{1-z}\right)_+
\label{eq:GammaFSR}
\eeqar
with $\gamma_E$ and $\Gamma(y)$ denoting Euler's constant and
the Gamma function, respectively, and the logarithms
\beq
\betamu = \frac{2\alpha(0)}{\pi} 
\left[\ln\biggl(\frac{Q^2}{m_\mu^2}\biggr)-1\right],
\qquad
\betae = \frac{2\alpha(0)}{\pi}\ln\biggl(\frac{Q^2}{\Me^2}\biggr).
\eeq
The parts solely proportional to a power of $\betamu$ correspond to
collinear (multi-)photon emission off the muon, the exponential 
factor describing resummed soft-photonic effects. The non-logarithmic
term ``$-1$'' in $\betamu$ accounts for a non-singular universal
soft-photonic correction.
The part proportional to $\betamu\betae$ corresponds to
collinear one-photon emission followed by a collinear splitting
$\gamma\to\Pep\Pem$. Treating this contribution as part of
$\Gamma_{\mu\mu}^{\LL}(z,Q^2)$ means that the event selection is
based on a muon momentum $z k_l$ also in this case, i.e.\ that
the $\Pep\Pem$ are considered as part of the hadronic and photonic
shower in the detector.
In this context one should also worry about the possibility of
$\mu^+\mu^-$ pairs in the final state resulting from photon splitting.
If one assumes that such muons escape the detector, i.e.\ are not part of
the hadronic shower, the corresponding
logarithmic contributions cancel, since the full $z$ range is
integrated over. 

Technically
we add the cross section \refeq{eq:llfsr} to the one-loop result and
subtract the lowest-order and one-loop contributions 
$\rd\sigma_{\CC,\LL^1\FSR}$ already contained 
within this formula, 
\beq
  \sigma_{\CC,\LL^1\FSR} =
  \int \rd\sigma_{\CC,0}(p_l,p_q;k_l,k_q) \int^1_0 \rd z \, 
  \left[\delta(1-z)+\Gamma_{\mu\mu}^{\LL,1}(z,Q^2)\right] \,
  \Theta_{\cut}(z k_l),
\eeq
in order to avoid double counting.
The one-loop contribution to the structure function reads
\beqar
  \Gamma_{\mu\mu}^{\LL,1}(z,Q^2) &=&
  \frac{\betamu}{4} \left(\frac{1+z^2}{1-z}\right)_+ .
\eeqar
More precisely, we adapt the value of $\alpha$ in
$\Gamma_{\mu\mu}^{\LL,1}(z,Q^2)$ to the chosen IPS, so that 
the $\alpha\ln(m_\mu^2/Q^2)$ contribution of the
${\cal O}(\alpha)$ correction is subtracted exactly.
Thus, the procedure of adding higher-order FSR effectively changes
also the value of $\alpha$ in the $\alpha\ln(m_\mu^2/Q^2)$ term
to $\alpha(0)$ which is the appropriate coupling for real-photonic
effects.

Finally,
note that the uncertainty that is connected with the choice of $Q^2$
enters in ${\cal O}(\alpha^2)$, since all $\Oa$ corrections,
including constant terms, are taken into account.
As default we choose the value
\beq
Q = \max\{\xi\sqrt{s},2m_\mu\}
\label{eq:FSR_scale}
\eeq
with $\xi=1$. In order to quantify the scale uncertainty, we vary $\xi$
between $0.3$ and $3$. The lower limit $2m_\mu$ accounts for the fact that 
the muon energy $\sqrt{s}/2$ in the CM frame cannot drop below $m_\mu$.

\section{Parton distribution functions}
\label{se:PDF}

\subsection{\boldmath{Inclusion of ${\cal O}(\alpha)$ corrections}}
\label{se:OaPDFs}

The inclusion of ${\cal O}(\alpha)$ corrections to hadronic cross
sections, which is described in the following in detail,
conceptually proceeds along the same lines as the incorporation of
next-to-leading (NLO) QCD corrections. 

\begin{sloppypar}
The ${\cal O}(\alpha)$-corrected parton cross sections
contain mass singularities of the form
$\alpha\ln(m_q)$, which are due to collinear photon radiation off the
initial-state quarks or due to a collinear splitting 
$\gamma\to q\bar q$ for initial-state photons.  
In complete analogy to factorization in NLO
QCD calculations, we absorb these collinear singularities into the
quark and photon distributions.  
For processes that involve only quarks and/or antiquarks in the initial state
in lowest order,
the factorization is achieved by replacing the (anti-)quark distribution
$q(x)$ according to 
\beqar
q(x) &\to& q(x,M^2) 
-\int_x^1 \frac{\rd z}{z} \, q\biggl(\frac{x}{z},M^2\biggr) \,
\frac{\alpha}{2\pi} \, Q_q^2 
\nn\\
&& \qquad {} \times
\biggl\{
\ln\biggl(\frac{M^2}{m_q^2}\biggr) \Bigl[ P_{ff}(z) \Bigr]_+
-\Bigl[ P_{ff}(z) \Bigl(2\ln(1-z)+1\Bigr) \Bigr]_+
+C_{ff}(z)
\biggr\}
\nn\\
&& {}
-\int_x^1 \frac{\rd z}{z} \, \gamma\biggl(\frac{x}{z},M^2\biggr)
\frac{\alpha}{2\pi} \, 3Q_q^2 
\biggl\{
\ln\biggl(\frac{M^2}{m_q^2}\biggr) P_{f\gamma}(z)
+C_{f\gamma}(z)
\biggr\},
\label{eq:factorization}
\eeqar
where $M$ is the factorization scale, $C_{ij}(z)$ are the so-called
coefficient functions, and the splitting functions $P_{ij}(z)$ are defined as
\beq
P_{ff}(z) = \frac{1+z^2}{1-z}, \qquad
P_{f\gamma}(z) = z^2+(1-z)^2.
\label{eq:splittings}
\eeq
The replacement \refeq{eq:factorization} defines the
same finite parts in the ${\cal O}(\alpha)$ correction, i.e.\ the same
coefficient functions, as the usual $D$-dimensional regularization
for exactly massless partons, where the $\ln(m_q)$ terms appear as
$1/(D-4)$ poles. We have derived this correspondence by explicit
calculation and found agreement with the results of \citere{Humpert:1980uv}.
The actual form of the coefficient functions
defines the finite parts of the ${\cal O}(\alpha)$ corrections and,
thus, the factorization scheme. 
Following standard definitions of QCD, we distinguish the
$\MSbar$ and DIS-like schemes which are formally defined by
\beqar
C^{\MSbar}_{ff}(z) &=& 
C^{\MSbar}_{f\gamma}(z) = 0,
\nn\\
C^{\DIS}_{ff}(z) &=& 
\left[ P_{ff}(z)\left(\ln\biggl(\frac{1-z}{z}\biggr)-\frac{3}{4}\right)
+\frac{9+5z}{4} \right]_+,
\nn\\
C^{\DIS}_{f\gamma}(z) &=& P_{f\gamma}(z)\ln\biggl(\frac{1-z}{z}\biggr)
-8z^2+8z-1.
\eeqar
The $\MSbar$ scheme is motivated by formal simplicity, because it
merely rearranges the UV-divergent terms (plus some trivial constants)
as defined in dimensional regularization.
The DIS-like scheme is defined in such a way that the DIS structure
function $F_2$ does not receive any corrections; in other words,
the radiative corrections to electron--proton DIS are implicitly 
contained in the PDFs in this case.
Whatever scheme has been adopted in the extraction of PDFs from 
experimental data, the same scheme has to be used when predictions
for other experiments are made using these PDFs.
\end{sloppypar}

The absorption of the collinear singularities of ${\cal O}(\alpha)$
into PDFs requires the
inclusion of the corresponding ${\cal O}(\alpha)$ corrections into the
Dok\-shitzer--Gri\-bov--Lipa\-tov--Alta\-relli--Pa\-risi
(DGLAP) evolution of these distributions and into their fit to
experimental data.
In the past, when the ${\cal O}(\alpha)$ effects were not yet included
in the PDF fit to data, the inclusion of the
${\cal O}(\alpha)$ corrections to the DGLAP evolution showed
\cite{Kripfganz:1988bd} that the impact of these corrections on the
quark distributions in the $\MSbar$ factorization scheme
should be smaller than remaining QCD uncertainties.
Therefore, ${\cal O}(\alpha)$ corrections to hadronic processes
were usually included in the $\MSbar$ factorization scheme using
the existing PDFs, which ignored ${\cal O}(\alpha)$ effects.
This was also done in our previous calculation \cite{Diener:2003ss}.

\begin{sloppypar}
Recently the MRST collaboration released the first set of PDFs 
\cite{Martin:2004dh}, called ``MRST2004QED'', 
that consistently include ${\cal O}(\alpha)$ corrections. 
The ${\cal O}(\alpha)$ effects induce two important modifications
in the PDFs. Firstly, the difference in the electric charges of the
up- and down-type quarks leads to a violation of the isospin
symmetry between proton and neutron PDFs. It was frequently 
pointed out in the literature 
\cite{Bernstein:2002sa,Gambino:2002xp,Londergan:2004nr,Zeller:2002du,%
McFarland:2003jw,Martin:2004dh}
that this isospin violation, i.e.\
$u^\Pp(x)\ne d^\Pn(x)$ etc., is of particular importance when
analyzing ratios of neutral-current to charged-current cross 
sections in neutrino DIS.
Secondly, the factorization of collinear singularities from
photon emission leads to photon distribution functions both for
the proton and the neutron. 
The corresponding photon-induced contributions to hadronic cross sections 
have not yet been considered in the literature (since the
photon PDF was not yet available), but have been assumed to be small,
because of an implicit suppression factor $\alpha$ in the radiatively
induced photon PDF. In our discussion of numerical results below
we illustrate the size of the ${\cal O}(\alpha)$ corrections to the PDFs
in neutrino DIS cross sections by switching these corrections on and off.
Moreover, we calculate and discuss the size of the photon-induced 
contributions to the
cross sections explicitly. Some technical details of this part of the
calculation are given in the Appendix.
\end{sloppypar}
 
The authors of the MRST2004QED set of PDFs did not explicitly state
in \citere{Martin:2004dh} which scheme is relevant for the factorization 
of the ${\cal O}(\alpha)$ corrections. They themselves did not include
any ${\cal O}(\alpha)$ corrections, neither in the underlying data set
nor in the used theoretical predictions, and used the same data as in
their standard fit, implicitly assuming that no ${\cal O}(\alpha)$ corrections
corresponding to photon emission off incoming quarks had been applied
in the data analysis. For the $F_2$ data analysis at HERA this
assumption was confirmed to us by members of the H1 and ZEUS
collaborations and by theorists involved in the analysis
\cite{privcom}.
Therefore, we conclude that the ${\cal O}(\alpha)$ corrections
induced by photon emission off incoming quarks implicitly went
into the PDFs in the fit, which thus precisely corresponds to the
DIS-like factorization. We use this scheme in the numerical analysis
below. 
We note, however, that the MRST2004QED PDFs are defined in NLO QCD
in the $\MSbar$ factorization scheme.

\subsection{Nuclear structure functions}

In reasonable approximation, nuclear structure functions can be
constructed from the PDFs for protons and neutrons upon appropriately
weighting the distribution for a specific parton according to the
probability of finding a proton or neutron in the target.
The NuTeV collaboration, e.g., quotes a neutron excess
of $\eps=(N-Z)/(N+Z)=5.74\%$ in their target, where $N$ and $Z$ 
are the (average) numbers of neutrons and protons in the target, respectively.
The probabilities $c_\Pp$ and $c_\Pn$ to find a proton or neutron, are thus
given by
\beq
c_\Pp = \frac{1}{2}(1-\eps), \qquad
c_\Pn = \frac{1}{2}(1+\eps).
\eeq
The corresponding nuclear structure functions read
\beq
f_a(x) = c_\Pp a^\Pp(x) + c_\Pn a^\Pn(x), \qquad a=q,\bar q,\gamma,
\nn\\
\eeq
where $a^\Pp(x)$ and $a^\Pn(x)$ generically denote the PDFs for the proton 
and neutron, respectively.
An isoscalar target obviously corresponds to the choice $\eps=0$, i.e.\
$c_\Pp = c_\Pn = \frac{1}{2}$.
In the following we make use of nuclear structure functions derived
with the value for the neutron excess $\eps$ quoted by NuTeV.

\section{Discussion of results}
\label{se:results}

\subsection{Input parameters and setup of the calculation}

For the numerical evaluation we use the following set of
SM parameters,
\beq\arraycolsep 3pt
\begin{array}[b]{lcllcllcl}
\GF & = & 1.16637 \times 10^{-5} \GeV^{-2}, \quad&
\alpha(0) &=& 1/137.03599911, &
\alpha(\MZ) &=& 1/128.93, \\
\MW & = & 80.425\GeV, &
\MZ & = & 91.1876\GeV, &
\alpha_{\mathrm{s}} &=& 0.1172,\\
\MH & = & 115\GeV, \\
\Me & = & 0.51099892\MeV, &
m_\mu &=& 105.658369\MeV,\quad &
m_\tau &=& 1.77699\GeV, \\
\Mu & = & 66\MeV, &
\Mc & = & 1.2\GeV, &
\Mt & = & 178\;\GeV, 
\\
\Md & = & 66\MeV, &
\Ms & = & 150\MeV, &
\Mb & = & 4.3\GeV, 
\end{array}
\label{eq:SMpar}
\eeq
which essentially follows \citere{Eidelman:2004wy}.  For the top-quark
mass $\Mt$ we have taken the value of
\citere{Azzi:2004rc}. The masses of the light quarks are adjusted to
reproduce the hadronic contribution to the photonic vacuum
polarization of \citere{Jegerlehner:2001ca}. 

\begin{sloppypar}
As explained in detail in \refse{se:OaPDFs}, we consistently use
the set of MRST2004QED PDFs introduced in \citere{Martin:2004dh},
which include ${\cal O}(\alpha)$ corrections. 
In order to quantify
the modifications induced by these corrections to the PDFs we
use the alternative set MRST2004QEDx, which we got from the MRST
collaboration and which differs from the original MRST2004QED set
only by neglecting the ${\cal O}(\alpha)$ effects.
Since we do not include NLO QCD corrections in our calculation,
it does not make sense to discuss the factorization scale dependence
of our results, because the QED and QCD factorization scales are
set equal in the MRST2004QED PDFs.
As already done in \citere{Diener:2003ss}, we set the QED
factorization scale $M^2$ to the momentum transfer 
of the leptonic line; the only difference comes from the fact that
the MRST2004QED PDFs are not delivered for $M$ values below their
start scale $M_0$. Therefore, we use 
\beq
M^2 = \max\left\{-(p_l-k_l)^2,M_0^2\right\}, \qquad M_0^2=1.25\GeV^2,
\eeq
where $p_l$ and $k_l$ are the incoming neutrino momentum and the
outgoing neutrino or lepton momentum, respectively.
Moreover, the PDFs require $x>10^{-5}$; otherwise they are set to zero.
\end{sloppypar}

The energy of the incoming neutrinos is set to
$E_\nu^{\LAB}=80\GeV$, but the energy of all outgoing hadrons and photons
is constrained by $E_{\had+\phot}^{\LAB} > 20\GeV$, which differs from
the setup used in \citere{Diener:2003ss}.

As done in \citeres{Bardin:1986bc,Diener:2003ss}, we quantify the
corrections to the $R^\nu$ ratio by
\beq
\delta R^\nu_\NC = \frac{\delta\sigma^\nu_\NC}{\sigma^\nu_\NC},
\qquad
\delta R^\nu_\CC = -\frac{\delta\sigma^\nu_\CC}{\sigma^\nu_\CC},
\eeq
where $\delta\sigma^\nu_{\NC/\CC}$ are the corrections to the NC and
CC cross sections. These corrections, to a reasonable approximation, 
translate into a shift in the extracted on-shell weak mixing angle
$\sin^2\theta_\PW$ according to \cite{Bardin:1986bc,Diener:2003ss}
\beq
\De\sin^2\theta_\PW =
\frac{\frac{1}{2}-\sin^2\theta_\PW+\frac{20}{27}\sin^4\theta_\PW}
{1-\frac{40}{27}\sin^2\theta_\PW}
\left( \delta R^\nu_\NC + \delta R^\nu_\CC \right).
\eeq

Technically we have extended our ${\cal O}(\alpha)$ calculation w.r.t.\ 
its description in \citere{Diener:2003ss} by implementing a
generalization of the dipole subtraction formalism to non-collinear-safe
observables.%
\footnote{The dipole subtraction formalism as described for real QCD
corrections \cite{Catani:2002hc} and photon radiation
\cite{Dittmaier:2000mb,Roth:1999kk} is originally set up for collinear-safe
observables where all final-state mass singularities cancel. 
The generalization to non-collinear-safe observables is partially described
(for charged final-state particles only) in \citere{Bredenstein:2005zk};
the complete generalization
will be published elsewhere \cite{dipoleinprep}.}
All numerical results below are obtained with this technique, which turns
out to be more efficient, but they have
been confirmed by phase-space slicing as described in \citere{Diener:2003ss}.

\subsection{Numerical results on integrated cross sections}

Table~\ref{tab:numres} summarizes our numerical results 
as obtained in the three different IPS ``$\alpha(0)$'', ``$\alpha(\MZ)$'',
and ``$\GF$''. 
\newcommand{\spc}{\phantom{-}}
\begin{table}
\centerline{
$\begin{array}{clllll}
\hline
\mbox{IPS} & \mbox{contribution}
& R^\nu_0 & \delta R^\nu_\NC & \delta R^\nu_\CC & \De\sin^2\theta_\PW
\\
\hline
\alpha(0)  & {\cal O}(\alpha), \mbox{PDFqed\phantom{x}} 
              & 0.30455 & \spc0.0562  & -0.0840  & -0.0130 \\
           & {\cal O}(\alpha), \mbox{PDFqedx} 
              & 0.31167 & \spc0.0555 & -0.0841 & -0.0134 \\
           & \mbox{univ.~2-loop}  & & \spc0.01275 & -0.01279 & -0.00002 \\
           & \mbox{univ.~2-loop}' & & \spc0.01229 & -0.01235 & -0.00003 \\
           & \mbox{h.o.~FSR}  & & \spc0 & \spc0.000014(1) & \spc0.000007(1) \\
           & \mbox{h.o.~FSR}' & & \spc0 & \spc0.000010  & \spc0.000005\\
           & \mbox{$\gamma$ process} & & -0.000005 & -0.000063 & -0.000032 \\
\cline{2-6}
           & \mbox{``best''} & & \spc0.0690 & -0.0968 & -0.0130 \\[.5em]
\hline
\alpha(\MZ)& {\cal O}(\alpha), \mbox{PDFqed\phantom{x}}
              & 0.30455 & -0.0660 & \spc0.0365 & -0.0138 \\
           & {\cal O}(\alpha), \mbox{PDFqedx} 
              & 0.31167 & -0.0668 & \spc0.0365 & -0.0142 \\
           & \mbox{univ.~2-loop}  & & \spc0.01248 & -0.01319 & -0.00033 \\
           & \mbox{univ.~2-loop}' & & \spc0.01258 & -0.01330 & -0.00033 \\
           & \mbox{h.o.~FSR}  & & \spc0 & \spc0.00073(1) & \spc0.00034(1)\\
           & \mbox{h.o.~FSR}' & & \spc0 & \spc0.00072 & \spc0.00034\\
           & \mbox{$\gamma$ process} & & -0.000005 & -0.000063 & -0.000032 \\
\cline{2-6}
           & \mbox{``best''} & & -0.0535 & \spc0.0240 & -0.0138 \\[.5em]
\hline
\GF        & {\cal O}(\alpha), \mbox{PDFqed\phantom{x}}
              & 0.30455 & \spc0.0016  & -0.0303  & -0.0135 \\
           & {\cal O}(\alpha), \mbox{PDFqedx} 
              & 0.31167 & \spc0.0009  & -0.0303 & -0.0138 \\
           & \mbox{univ.~2-loop} & & -0.00039 & \spc0 & -0.00018 \\
           & \mbox{univ.~2-loop}' & & -0.00039 & \spc0 & -0.00018 \\
           & \mbox{h.o.~FSR}  & & \spc0 & \spc0.00040(1) & \spc0.00019(1) \\
           & \mbox{h.o.~FSR}' & & \spc0 & \spc0.00039 & \spc0.00018 \\
           & \mbox{$\gamma$ process} & & -0.000005 & -0.000063 & -0.000032 \\
\cline{2-6}
           & \mbox{``best''} & & \spc0.0012 & -0.0299 & -0.0135 \\
\hline
\end{array}$ }
\caption{Results on the ratio $R^\nu$ with 
$E_\nu^{\LAB}=80\GeV$, $E_{\had+\phot}^{\LAB} > 20\GeV$
in lowest order ($R^\nu_0$),
corrections from NC and CC cross sections
($\delta R^\nu_\NC$ and $\delta R^\nu_\CC$),
and shift $\De\sin^2\theta_\PW$, for various input-parameter schemes (IPS).
All but the ``PDFqedx'' numbers are obtained with the 
${\cal O}(\alpha)$-corrected PDFs MRST2004QED; in PDFqedx the
${\cal O}(\alpha)$ corrections within the PDFs are switched off.
The different rows correspond to the pure ``${\cal O}(\alpha)$'' corrections,
to improvements by ``universal 2-loop'' corrections, to
``higher-order FSR'' corrections, and to contributions from the
``$\gamma$ process'' with a photon in the initial state. 
The number in parentheses indicates the statistical integration error
in the last digit.
For more details, see main text. 
}
\label{tab:numres}
\end{table}
In the first row, denoted ``${\cal O}(\alpha), \mbox{PDFqed}$'',
we give the cross-section ratio $R^\nu_0$ in lowest order
as well as the corrections
$\delta R^\nu_\NC$, $\delta R^\nu_\CC$, $\De\sin^2\theta_\PW$ based on
${\cal O}(\alpha)$ corrections only. The results are based on the
${\cal O}(\alpha)$-corrected PDFs MRST2004QED \cite{Martin:2004dh}, but
contributions from the $\gamma$-induced processes are not taken into account
in this row. 
This means, apart from the different kinematical setup, the new PDFs
(with DIS-like QED factorization), and
the new input parameters, the first row is based on the calculation
of ${\cal O}(\alpha)$ corrections described in \citere{Diener:2003ss}. 
The second row, denoted ``${\cal O}(\alpha), \mbox{PDFqedx}$'', 
contains the corresponding results obtained with the set of PDFs
MRST2004QEDx that differ from MRST2004QED only in the consistent
neglect of the ${\cal O}(\alpha)$ corrections in the PDFs.
Comparing the first two rows for each IPS, the influence of the
implicit ${\cal O}(\alpha)$ corrections in the PDFs can be quantified
to about $0.0004$ in $\sin^2\theta_\PW$.
All the remaining rows in \refta{tab:numres} are calculated with the
${\cal O}(\alpha)$-corrected PDFs MRST2004QED.

The rows denoted ``univ.~2-loop'' and ``univ.~2-loop$^\prime\,$''
show the impact of the universal electroweak corrections  beyond 
${\cal O}(\alpha)$ induced by $\De\alpha$ and $\Delta\rho$, as
worked out in \refse{se:llho}. In detail, the first of these rows
correspond to the resummed versions of the two-loop effects
[first equations in
\refeq{eq:cc_llho_alp0}--\refeq{eq:cc_llho_gf} and
\refeq{eq:nc_llho_alp0}--\refeq{eq:nc_llho_gf}],
while the second corresponds to the genuine two-loop parts
[second equations in \refeq{eq:cc_llho_alp0} etc.].
The difference between the two versions is $\lsim 10^{-5}$ in
$\sin^2\theta_\PW$ and, thus, negligible.
Note that the size of these universal two-loop corrections to the
cross sections is rather large, more than $1\%$, in the $\alpha(0)$
and $\alpha(\MZ)$ schemes. 
In the $\GF$ scheme the universal effects beyond ${\cal O}(\alpha)$ 
are very small also for the individual cross sections, reflecting
the fact that they are widely absorbed in the lowest order.
Since the effects enter CC and NC cross sections in a similar way,
they widely cancel in the cross-section ratio and influence
$\sin^2\theta_\PW$ only at the level of $0.0003$, depending on the IPS.

\begin{sloppypar}
The result on higher-order FSR corrections to the CC cross section are 
indicated by ``h.o.~FSR'' and ``h.o.~FSR$^\prime\,$'' in \refta{tab:numres},
where the first of these rows result from the structure function as given
in \refeq{eq:GammaFSR} and the second from an expansion of \refeq{eq:GammaFSR}
up to ${\cal O}(\alpha^3)$. The differences between these numbers, which
quantify leading logarithmic FSR effects beyond ${\cal O}(\alpha^3)$, are 
negligible. The corrections in the $\alpha(\MZ)$- and $\GF$-schemes
are significantly larger than in the $\alpha(0)$-scheme, because
in those schemes the higher-order FSR corrections also
receive a contribution from the ${\cal O}(\alpha)$ part, more
precisely from the ${\cal O}(\alpha)$ part calculated with $\alpha(0)$
minus the ${\cal O}(\alpha)$ part calculated with the
$\alpha$ from the chosen IPS.
In fact almost the entire effect is due to this change from $\alpha$
to $\alpha(0)$ in the ${\cal O}(\alpha)$ FSR, which is $1.14\%$
in $\delta R^\nu_\CC$ when calculated with $\alpha(0)$.
In the $\GF$ scheme, higher-order FSR affects $\sin^2\theta_\PW$ at 
the level of $0.0002$.

As explained in \refse{se:llfsr}, our results depend on the QED
splitting scale $Q$ for collinear FSR in ${\cal O}(\alpha^2)$ and beyond,
while in ${\cal O}(\alpha)$ the scale $Q$ is fixed by the non-logarithmic 
corrections.
Table~\ref{tab:fsr_scale} illustrates the residual scale dependence
by varying the parameter $\xi$ in \refeq{eq:FSR_scale} between $0.3$
and $3$; the default choice is $\xi=1$.
\begin{table}
\centerline{
$\begin{array}{clcll}
\hline
\mbox{IPS} & \mbox{contribution} &
\xi & \delta R^\nu_\CC & \De\sin^2\theta_\PW
\\
\hline
\alpha(0)  & \mbox{h.o.~FSR}' & 0.3 &    -0.000008  &    -0.000004\\
           &                  & 1   & \spc0.000010  & \spc0.000005\\
           &                  & 3   & \spc0.000040  & \spc0.000019\\
\hline
\alpha(\MZ)& \mbox{h.o.~FSR}' & 0.3 & \spc0.00046 & \spc0.00021\\
           &                  & 1   & \spc0.00072 & \spc0.00034\\
           &                  & 3   & \spc0.00099 & \spc0.00046\\
\hline
\GF        & \mbox{h.o.~FSR}' & 0.3 & \spc0.00024 & \spc0.00011 \\
           &                  & 1   & \spc0.00039 & \spc0.00018 \\
           &                  & 3   & \spc0.00055 & \spc0.00026 \\
\hline
\end{array}$ }
\caption{Scale dependence of higher-order FSR corrections.}
\label{tab:fsr_scale}
\end{table}
The table shows that in the $\alpha(0)$-scheme 
the scale uncertainty is of the order of $0.00002$
when translated into a shift in $\sin^2\theta_\PW$.
While this number corresponds to genuine FSR effects beyond 
${\cal O}(\alpha)$, the larger scale dependence in the $\alpha(\MZ)$- and
$\GF$-schemes originates from the scale change in the term
$[\alpha(0)/\pi-\alpha/\pi]\ln(m_\mu^2/Q^2)$, which changes the coupling $\alpha$ to
$\alpha(0)$ in the ${\cal O}(\alpha)$ FSR corrections.
\end{sloppypar}

The contributions of the $\ga$-induced processes are shown in
\refta{tab:numres} in the rows ``$\ga$~process''.
Since we have set the photon coupling to $\alpha(0)$ in all three IPS,
the relative impact of these corrections is identical in all the IPS.
The effect on $\sin^2\theta_\PW$ is about $0.00003$ and, thus, 
negligible.

The numbers called ``best'' in \refta{tab:numres}
show the sum of the ${\cal O}(\alpha)$ and higher-order corrections,
where the latter are the sum of the contributions labelled
``univ.~2-loop'', ``h.o.~FSR$'$'', and 
``$\gamma$~process''.
The fact that these ``best'' numbers agree with the
purely ${\cal O}(\alpha)$-corrected results up to $\lsim 0.0001$ in
$\sin^2\theta_\PW$ is accidental, since universal effects and
FSR, which widely cancel each other, are of different physical origin.
This means, in particular, the cancellation will not take place for
a different kinematical setup (neutrino energy, hadronic energy cut, etc.). 

Finally, in \refta{tab:TU} we estimate the theoretical uncertainties 
from missing electroweak higher-order effects from the results discussed above.
\begin{table}
\centerline{
$\begin{array}{ll}
\hline
\mbox{source} & \De\sin^2\theta_\PW
\\
\hline
{\cal O}(\alpha) \mbox{ effect in PDFs} & 0.0004 \\
\hline
\mbox{univ.\ electroweak effects} & 0.00003 \\
\mbox{h.o.~FSR}  & 0.00002 \\
\mbox{non-univ.\ electroweak effects} & 0.0003\\
\hline
\end{array}$ }
\caption{Estimates of theoretical uncertainties from missing
electroweak higher-order effects in the $\GF$-scheme.}
\label{tab:TU}
\end{table}
Table~\ref{tab:numres} clearly shows that the ${\cal O}(\alpha)$ corrections
implicit in the PDFs affects $\sin^2\theta_\PW$ of the order $0.0004$,
which is not negligible w.r.t.\ the error of about $0.0016$ quoted by
NuTeV. ${\cal O}(\alpha)$-corrected PDFs should therefore be used in such
an analysis. If this is done, the error estimate of $0.0004$ will
of course be reduced.
In \refta{tab:TU} we estimate the missing universal electroweak effects
in the $\GF$-scheme, which is the most reliable IPS, by changing the
parameter $\De\rho$ by 2\% of its size, which is of the order of effects 
beyond two loops \cite{Faisst:2003px}. The resulting uncertainty is
about $0.00003$ and thus negligible.
Missing higher-order effects from FSR have been quantified to be of the
order $0.00002$ in \refta{tab:fsr_scale} above. 
We take this value from the $\alpha(0)$-scheme, because the larger
uncertainties from changing $\alpha$ to $\alpha(0)$ in the two 
other schemes are part of non-universal electroweak effects 
beyond ${\cal O}(\alpha)$.
These non-universal electroweak effects 
are estimated as follows. In the $\GF$-scheme, the ${\cal O}(\alpha)$
corrections to the $R^\nu$ ratio are about $2.9\%$ where about
$1.1\%$ are due to FSR ($\xi=1)$. 
The relative coupling of FSR is clearly
$\alpha(0)$, but the effective value of $\alpha$ in the non-universal
$1.8\%$ can only be fixed in a two-loop calculation. Therefore,
changing the value of $\alpha_{\GF}$ to $\alpha(0)$ or $\alpha(\MZ)$,
i.e.\ varying the $1.8\%$ by about $3\%$, should give an idea about
missing electroweak two-loop effects. This procedure changes
$\sin^2\theta_\PW$ by $0.0003$, which means that missing non-universal
electroweak effects should dominate the theoretical uncertainty from
electroweak corrections.

\subsection{Numerical results on differential cross sections}

\begin{figure}
{\unitlength 1cm
\begin{picture}(16,8)
\put(-4.9,-17.0){\includegraphics{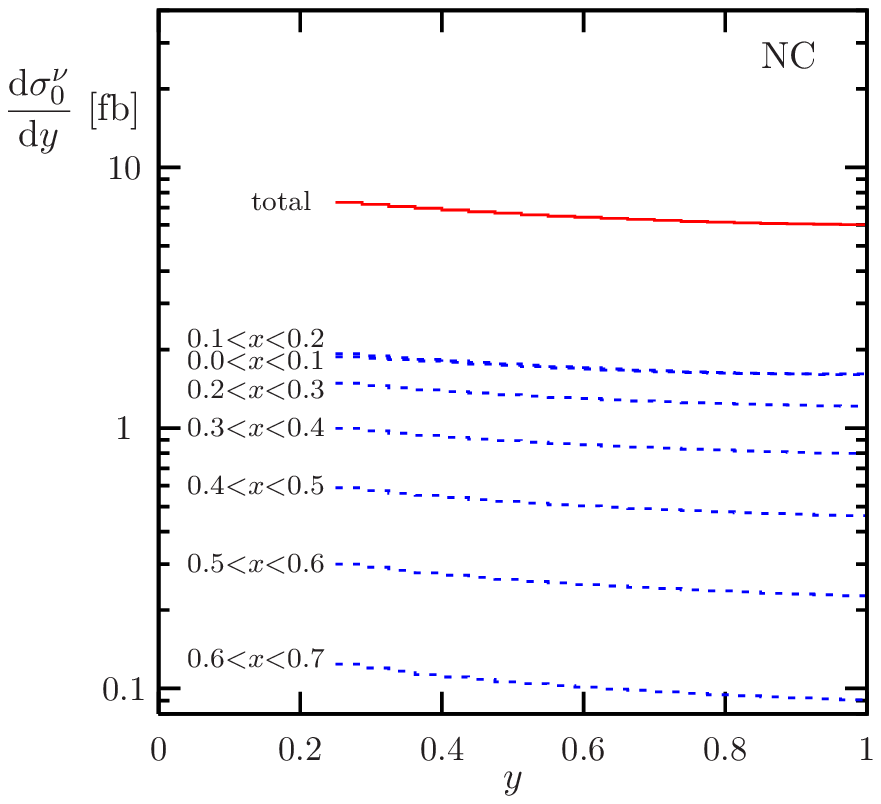}}
\put( 2.7,-17.0){\includegraphics{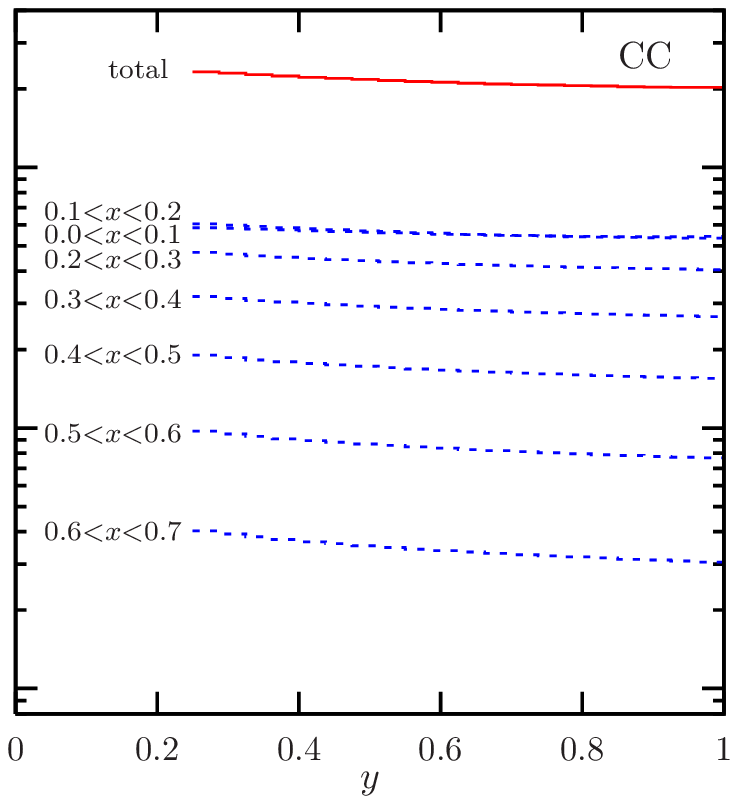}}
\end{picture}}%
\caption{Lowest-order predictions for the distributions in $y$ (total)
and individual contributions from different ranges in $x$ ($=x_{\mathrm{Bj}}$
in lowest order). The NC channel for $\nu$ scattering is shown on the l.h.s.,
the CC channel on the r.h.s.}
\label{fi:born}
\vspace*{2em}
{\unitlength 1cm
\begin{picture}(16,8)
\put(-4.9,-17.0){\includegraphics{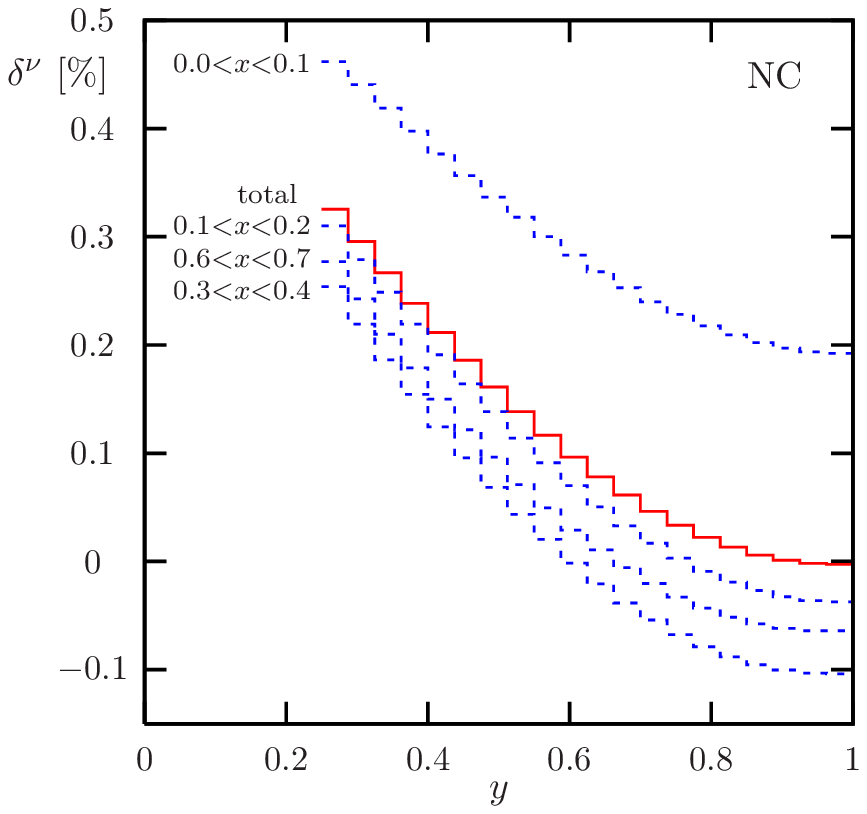}}
\put( 2.7,-17.0){\includegraphics{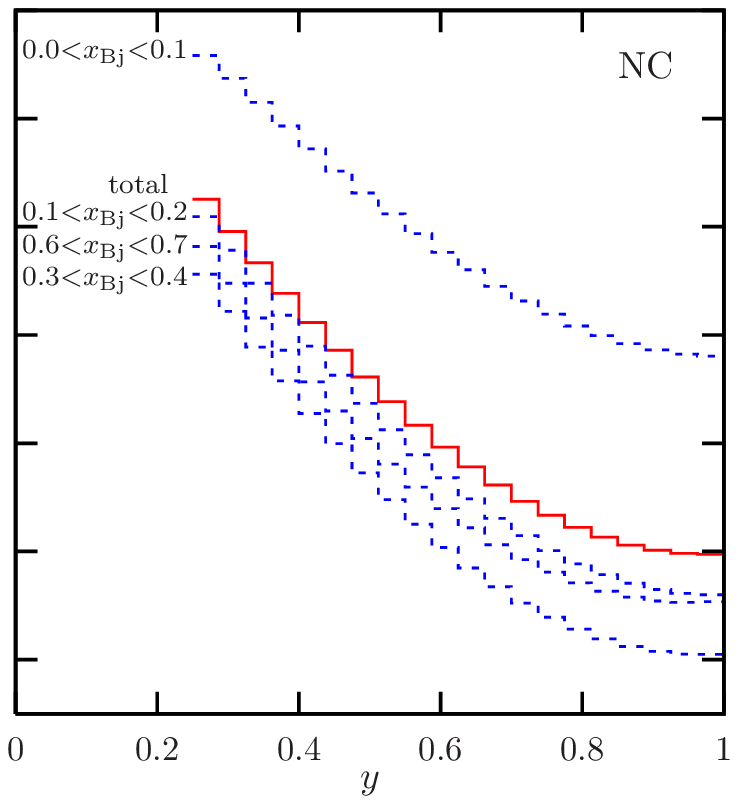}}
\end{picture}}%
\caption{Relative electroweak corrections 
$\delta^\nu=\delta\sigma^\nu/\sigma^\nu_0$
to the distributions in $y$ (total)
and to individual contributions from different ranges in $x$ (l.h.s.) and
in $x_{\mathrm{Bj}}$ (r.h.s.).}
\label{fi:corrnc}
\end{figure}

\begin{figure}
{\unitlength 1cm
\begin{picture}(16,16)
\put(-4.9,- 8.7){\includegraphics{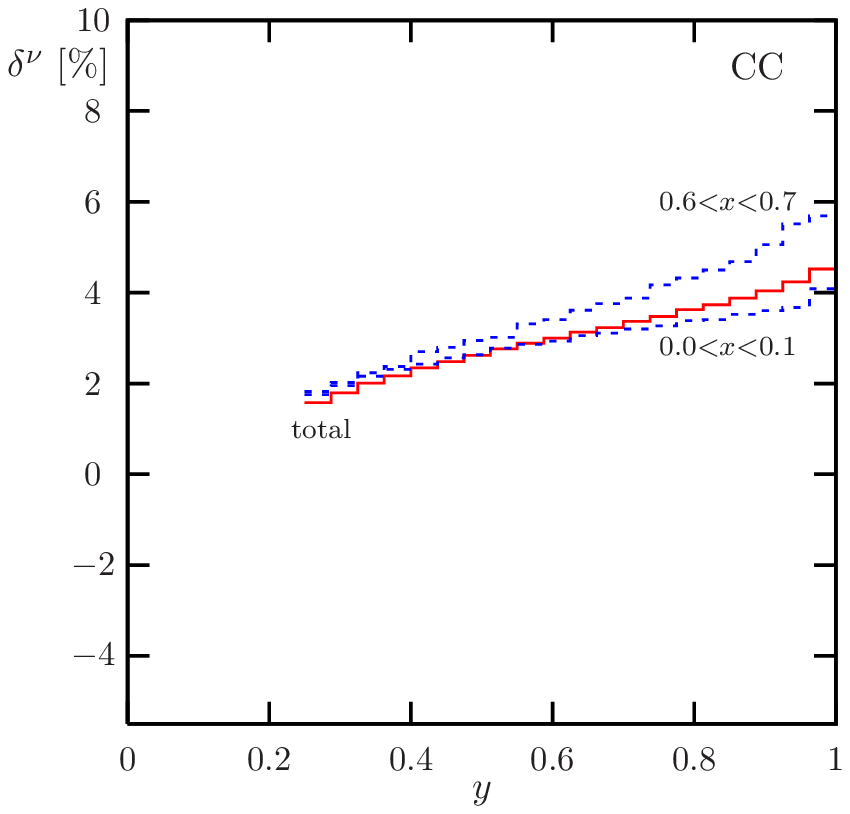}}
\put( 2.7,- 8.7){\includegraphics{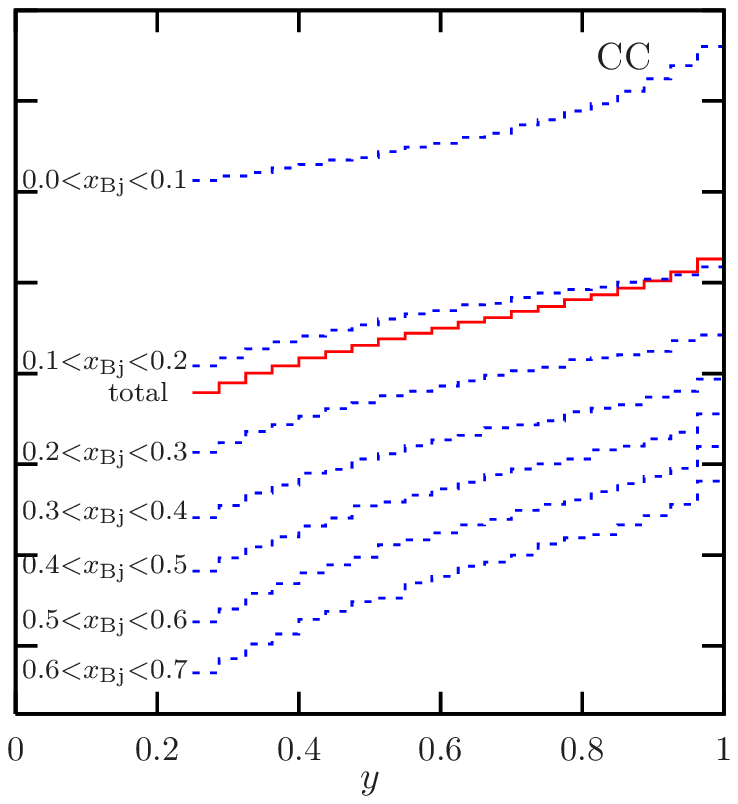}}
\put(-4.9,-17.0){\includegraphics{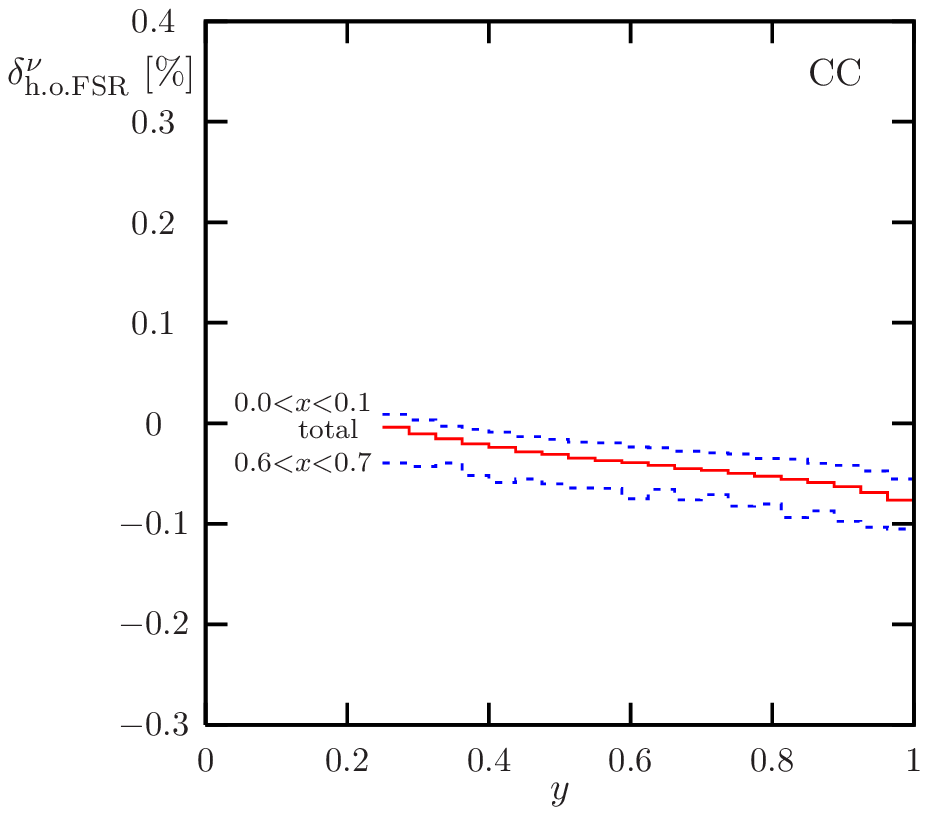}}
\put( 2.7,-17.0){\includegraphics{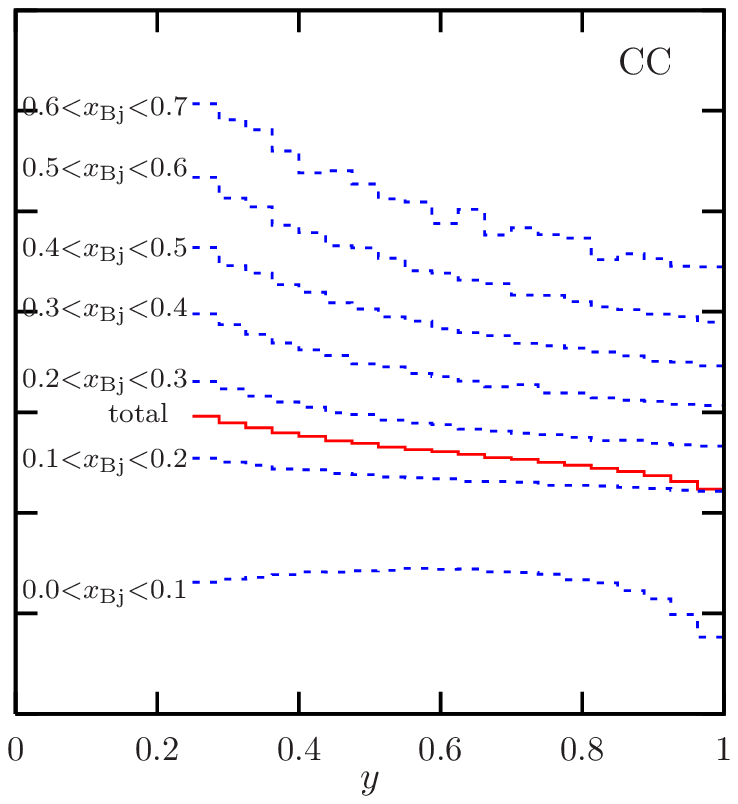}}
\end{picture}}%
\caption{Relative electroweak corrections 
$\delta^\nu=\delta\sigma^\nu/\sigma^\nu_0$
(upper plots) and higher-order FSR
(lower plots) to the distributions in $y$ (total)
and to individual contributions from different ranges in $x$ (l.h.s.) and
in $x_{\mathrm{Bj}}$ (r.h.s.).}
\label{fi:corrcc}
\end{figure}

Figures \ref{fi:born}--\ref{fi:corrcc} illustrate the differential
cross sections $\rd\si/\rd y$ and $\rd^2\si/\rd x/\rd y$ and the
corresponding corrections.
The curves labelled ``total'' correspond to $\rd\si/\rd y$ shown as
histograms, the other curves correspond to $\rd^2\si/\rd x/\rd y$
integrated over the indicated $x$ range.
Note that we actually distinguish between the variable $x$,
which is the argument of the PDFs, and the Bjorken variable $\xbj$,
which is reconstructed from initial- and final-state variables.
The lowest-order predictions of \reffi{fi:born} show that the integrated
cross sections, both for the NC and CC channel, are dominated by the
$x$ range ($x=\xbj$ in lowest order) $x\lsim 0.4$, while the
distributions in $y$ are flat.
The corrections to the NC process (\reffi{fi:corrnc}) vary between
$-0.1\%$ and $+0.5\%$ for all $x$, $\xbj$, and $y$ values. The distributions
in $x$ and $\xbj$ are almost identical, and the corrections in
the various $x$ and $\xbj$ ranges do not deviate by more than $0.2\%$
from the correction to the corresponding fully integrated quantity 
$\rd\si/\rd y$. 
The situation is very different in the CC case, as shown in \reffi{fi:corrcc}.
The corrections to $\rd\si/\rd y$ vary within $\sim2{-}4\%$.
The upper left plot in \reffi{fi:corrcc} demonstrates that the
corrections vary very little in $x$, but the upper right plot shows
a very large variation in $\xbj$.
This is due to the fact that binning w.r.t.\ the parton $x$ does not
destroy the inclusiveness of collinear FSR off the muon, but 
binning in $\xbj$ does.
In other words, a cut in $x$ does not cut into the cone of collinear FSR,
but a cut in $\xbj$ does.
These features also show up in higher-order FSR as depicted in the lower
plots of \reffi{fi:corrcc}.

The results on the distributions are particularly interesting for the
issue of implementing electroweak corrections in Monte Carlo generators
upon reweigthing certain distributions. The corrections to the differential
CC cross sections suggest that a reweighting in distributions w.r.t.\
$x$ is less delicate than in $\xbj$.

\section{Conclusions}
\label{se:concl}

A thorough understanding of electroweak corrections to 
deep-inelastic neutrino scattering is indispensable for exploiting
measurements of NuTeV and NOMAD to extract a value for the
on-shell weak mixing angle $\sin^2\theta_\PW=1-\MW^2/\MZ^2$.
In this paper, we have supplemented our previous calculation of 
electroweak ${\cal O}(\alpha)$ corrections by higher-order effects,
in order to reduce the theoretical uncertainty from unknown
electroweak corrections.

We take into account universal two-loop effects from 
$\De\alpha$ and $\De\rho$ as well as higher-order final-state photon
radiation off muons in the structure function approach.
The impact on $\sin^2\theta_\PW$ of these effects are about $0.0003$;
compared to the experimental uncertainty quoted by NuTeV, which is
about $0.0016$, these effects are small.

\begin{sloppypar}
Moreover, we employ the recently released ${\cal O}(\alpha)$-improved
parton distributions MRST2004QED and identify the relevant QED
factorization scheme, which is DIS like.
Switching on and off the implicit ${\cal O}(\alpha)$ corrections in
the parton distributions, changes $\sin^2\theta_\PW$ at the level of
$0.0004$.
\end{sloppypar}

As a technical byproduct, we describe two variants of phase-space
slicing and the dipole subtraction method for calculating processes
with initial-state photons. Such photon-induced processes become
relevant, because the ${\cal O}(\alpha)$ evolution of the parton 
distributions generates a photon distribution. 
Owing to the smallness of the available scattering energy, the impact
of photon-induced processes on $\sin^2\theta_\PW$ is negligible
for NuTeV.

A numerical discussion of the higher-order effects suggests that the
remaining theoretical uncertainty from unknown electroweak 
corrections is dominated by unknown non-universal two-loop effects
and is of the order 0.0003 when translated into a shift in
$\sin^2\theta_\PW$. 


\section*{Acknowledgments}

We would like to thank the MRST collaboration for delivering
a modified version of the PDFs of \citere{Martin:2004dh} in which the 
${\cal O}(\alpha)$ are switched off.
Moreover, we gratefully acknowledge V.~Lendermann, H.~Spiesberger, 
J.~Stirling and A.~Tapper for valuable discussions about the
treatment of photon radiation off quarks in the analysis of
DIS data and its implication for the use of the new MRST PDFs.
This work was supported in part by INTAS 03-51-4007.


\appendix
\section*{Appendix}
\renewcommand{\theequation}{A.\arabic{equation}}

\section*{Calculation of photon-induced cross sections}

As explained in \refse{se:OaPDFs} in detail, a consistent evaluation
of ${\cal O}(\alpha)$ corrections to quark-induced hadronic 
processes includes an evaluation of related photon-induced channels.
In neutrino DIS we have the leading-order processes
$\nu_\mu q \to \nu_\mu q$ and $\nu_\mu q \to \mu^- q'$, with
$q$ and $q'$ generically denoting the relevant quarks and antiquarks.
The corresponding photon-induced channels are
$\nu_\mu \gamma \to \nu_\mu q \bar q$ and
$\nu_\mu \gamma \to \mu^- q' \bar q$.
The matrix elements for these reactions obviously follow upon crossing
from the bremsstrahlung processes
$\nu_\mu q \to \nu_\mu q\gamma$ and $\nu_\mu q \to \mu^- q'\gamma$,
which already entered the calculation of ${\cal O}(\alpha)$ corrections
described in \citere{Diener:2003ss}. Alternatively they can also be
derived upon crossing the results for the real ${\cal O}(\alpha)$ corrections
to Drell--Yan-like W~production as explicitly given in 
\citere {Dittmaier:2001ay}.
We set the coupling factors such that the squared matrix element is
proportional to $\alpha^2\alpha(0)$, where $\alpha$ is equal to
$\alpha(0)$, $\alpha(\MZ)$, or $\alpha_{\GF}$ according to the respective
IPS; the factor $\alpha(0)$ is taken independent of the IPS, because this
is the correct effective coupling for the incoming photon which is
on shell.
Since the matrix elements have been derived for massless fermions the
integration over the domains of collinear initial-state splittings
$\gamma\to f\bar f$ requires a restoration of finite-fermion-mass effects.
The resulting mass-singular logarithms of initial-state quarks are
absorbed into the photon PDF as described in \refse{se:OaPDFs}.

In the following we describe three simple methods to restore such
effects in a generic reaction $a(p_a)+\gamma(k) \to f(p_f)+X$, 
where $a$ stands for
any massless incoming parton and $f$ is the light fermion or antifermion
whose mass effects we are interested in. The momenta of the particles are 
given in the parentheses. The remainder $X$ may contain
additional light fermions which can be treated in the same way as $f$.
The results obtained with the three methods are in mutual numerical
agreement, where the dipole subtraction method yields the smallest 
integration error when using the same statistics in all the methods.

\paragraph{Effective collinear factor}

The collinear singularity in the squared matrix element
$|\M_{a\gamma\to fX}|^2$ occurs if the angle $\theta_f$ between $f$ and
the incoming $\gamma$ becomes small; in this limit the scalar product
$(kp_f)$ is of ${\cal O}(m_f^2)$, where $m_f$ is the small mass of $f$.
Neglecting terms that vanish in the limit $m_f\to0$ the polarization-summed
squared matrix element 
(including the average over initial-state polarizations)
asymptotically behaves like \cite{Catani:2002hc}
\beq
\sum_{\mathrm{pol}} |\M_{a\gamma\to fX}(p_a,k)|^2 \;\asymp{kp_f\to0}\;
\frac{Q_f^2e^2}{x(kp_f)}\left[1-2x(1-x)+\frac{m_f^2}{kp_f}x\right]
\sum_{\mathrm{pol}} |\M_{a\bar f\to X}(p_a,p_{\bar f}=xk)|^2,
\label{eq:aff-fact}
\eeq
where $x=1-p_f^0/k^0$, $Q_fe$ is the electric charge of $f$,
and $\M_{a\bar f\to X}$ is the matrix element of the related process
which results from $a\gamma(\to af\bar f)\to fX$ upon cutting the $\bar f$ line
if this process proceeds in the collinear limit mainly via the 
$\gamma\to f\bar f$ splitting. The incoming momenta relevant in the
different matrix elements are given in parentheses.

The factorization formula \refeq{eq:aff-fact} can be used to relate
the two versions for the squared matrix element 
$|\M_{a\gamma\to fX}|^2$ obtained with a finite or vanishing fermion mass
$m_f$, just by eliminating $|\M_{a\bar f\to X}|^2$.
The result is
\beq
\sum_{\mathrm{pol}} |\M_{a\gamma\to fX}(p_a,k)|^2 \;\asymp{kp_f\to0}\;
g(m_f,x,k^0,\theta_f) \,
\sum_{\mathrm{pol}} |\M_{a\gamma\to fX}(p_a,k)|^2\Big|_{m_f=0}
\label{eq:m2ecf}
\eeq
with the ``effective collinear factor''
\beqar
\lefteqn{g(m_f,x,k^0,\theta_f)}
\nn\\
&=& \frac{(1-x)^2}{(1-x)^2+x^2} \,
\frac{4(k^0)^2\sin^2\frac{\theta_f}{2}
  \left[m_f^2+4(k^0)^2\sin^2\frac{\theta_f}{2}(1-x)^2
	\left((1-x)^2+x^2\right)\right]}
{\left[m_f^2+4(k^0)^2\sin^2\frac{\theta_f}{2}(1-x)^2\right]^2}.
\hspace{2em}
\eeqar
The squared matrix element on the l.h.s.\ of Eq.~\refeq{eq:m2ecf}
can be integrated over the phase space for a massless $f$, yielding
the correct result for a massive $f$ up to terms that are suppressed
by powers of $m_f$. This means that the collinear singularity is
regularized by the correct cutoff term $\ln m_f$ supplemented with
the correct constant contributions for $m_f\to0$. However, there are
spurios terms of ${\cal O}(m_f/Q)$ where $Q$ is a typical scale in the
process. 

\paragraph{Phase-space slicing}

In the collinear limit $\theta_f\to0$, $m_f\to0$ the phase-space volume element
$\rd\Phi_{a\gamma\to fX}(p_a,k)$ can be factorized as
\beq
\int_{\theta_f<\Delta\theta} \rd\Phi_{a\gamma\to fX}(p_a,k) 
\;\asymp{\Delta\theta\to0}\;
\int_0^1\rd x\, \int \rd\Phi_{a\bar f\to X}(p_a,xk) \,
\frac{(1-x)(k^0)^2}{2(2\pi)^2} \int_0^{\Delta\theta}\rd\theta_f \,\sin\theta_f,
\eeq
where $\Delta\theta\ll1$ is a small cutoff parameter defining the collinear
region. Here we also used that the incoming parton $a$ is massless.
Making use of this factorized phase space and of the asymptotic
form \refeq{eq:aff-fact} of the squared matrix element, the integral
over small emission angles $\theta_f$ can be carried out, yielding
\beqar
\sigma_{a\gamma\to fX}(p_a,k)\Big|_{\theta_f<\Delta\theta} &=&
\frac{1}{4p_a k}\int_{\theta_f<\Delta\theta} \rd\Phi_{a\gamma\to fX}(p_a,k) \,
\sum_{\mathrm{pol}} |\M_{a\gamma\to fX}(p_a,k)|^2
\nn\\
&=& \frac{Q_f^2\alpha}{2\pi}
\int_0^1\rd x\, H(m_f,x,k^0,\Delta\theta) \,
\frac{1}{4(p_a k)x}
\int \rd\Phi_{a\bar f\to X}(p_a,xk) \,
\nn\\
&& \qquad \times {}
\sum_{\mathrm{pol}} |\M_{a\bar f\to X}(p_a,p_{\bar f}=xk)|^2
\;+\; {\cal O}(\Delta\theta),
\nn\\
&=& N_{\mathrm{c},f}\,
\frac{Q_f^2\alpha}{2\pi}\int_0^1\rd x\, H(m_f,x,k^0,\Delta\theta) \,
\sigma_{a\bar f\to X}(p_a,p_{\bar f}=xk) \;+\; {\cal O}(\Delta\theta)
\nn\\          
\eeqar
with the auxiliary function
\beq
H(m_f,x,k^0,\Delta\theta) = P_{f\gamma}(x) 
  \ln\biggl(\frac{\Delta\theta^2(k^0)^2(1-x)^2}{m_f^2}\biggr) +2x(1-x) 
\eeq
containing the collinear singularity. The splitting function
$P_{f\gamma}(x)$ is defined in Eq.~\refeq{eq:splittings}, 
and $N_{\mathrm{c},f}$ is the colour factor for the fermion $f$
($N_{\mathrm{c,lepton}}=1$, $N_{\mathrm{c,quark}}=3$).

The remaining phase-space integration over the non-collinear region
$\theta_f>\Delta\theta$ can be carried out for a massless $f$
(provided no other singularities are connected with $f$)
without modifications.

\paragraph{Dipole subtraction}

The general idea of a subtraction method is to subtract and
to add a simple auxiliary function from the singular integrand.
This auxiliary function has to be chosen such that it cancels all
singularities of the original integrand so that the phase-space
integration of the difference can be performed numerically.
Moreover, the auxiliary function has to be simple enough so that it can
be integrated over the singular regions analytically, when the
subtracted contribution is added again. 
The dipole subtraction method, originally worked out for massless QCD
\cite{Catani:1996jh}, provides a general algorithm for the construction 
of this auxiliary function and for its integrated counterpart. 

The dipole subtraction method for photon radiation off massive
and massless fermions has been described in \citere{Dittmaier:2000mb}
and \citere{Roth:1999kk}. Following the same strategy, we work out
a similar formalism for the collinear splitting $\gamma\to f\bar f$
of a photon in the initial state into a light $f\bar f$ pair.
More generalizations of the dipole formalism of \citere{Dittmaier:2000mb}
will be published elsewhere \cite{dipoleinprep}.

The function that is subtracted from the integrand
$\sum_{\mathrm{pol}} |\M_{a\gamma\to fX}(p_a,k)|^2$,which is derived for
$m_f=0$, is defined as follows,%
\footnote{This construction closely follows the case of an 
``initial-state emitter'' and of an ``initial-state spectator'' 
in the case of photon or gluon radiation, as described in 
\citeres{Catani:1996jh,Dittmaier:2000mb,Roth:1999kk}.}
\beq
|\M_{\sub}|^2 = Q_f^2 e^2 \, h_{\gamma a}(k,p_f,p_a) \,
\sum_{\mathrm{pol}} 
|\M_{a\bar f\to X}(p_a,p_{\bar f}=x_{\gamma a}k,\tilde k_X)|^2,
\label{eq:Msub}
\eeq
with the radiator function
\beq
h_{\gamma a}(k,p_f,p_a) = 
\frac{P_{f\gamma}(x_{\gamma a})}{x_{\gamma a}(p_f k)}, \qquad
x_{\gamma a} = \frac{p_a k-p_f k-p_a p_f}{p_a k}.
\eeq
The final-state momenta entering the squared matrix element on the r.h.s.\
of Eq.~\refeq{eq:Msub}, generically denoted $\tilde k_X$, follow from the
original momenta $k_X$ upon the Lorentz transformation
\beq
\tilde k_X^\mu = {\Lambda^\mu}_\nu k_X^\nu
\eeq
with
\beq
{\Lambda^\mu}_\nu =
{g^\mu}_\nu - \frac{(P+\tilde P)^\mu(P+\tilde P)_\nu}{P^2+P\tilde P}
+\frac{2\tilde P^\mu P_\nu}{P^2}, \qquad
P^\mu=p_a^\mu+k^\mu-p_f^\mu, \quad \tilde P^\mu=p_a^\mu+x_{\gamma a}k^\mu.
\eeq
It is straightforward to check that $|\M_{\sub}|^2$ possesses the same
asymptotic behaviour as $\sum_{\mathrm{pol}} |\M_{a\gamma\to fX}|^2$
in Eq.~\refeq{eq:aff-fact} for $m_f=0$.
Thus, the difference $\sum_{\mathrm{pol}} |\M_{a\gamma\to fX}|^2-|\M_{\sub}|^2$
can be integrated numerically for $m_f=0$.
The correct dependence of $|\M_{\sub}|^2$ (and the related kinematics)
on a finite $m_f$ is, however, needed when this function is integrated
over $\theta_f$ leading to the collinear singularity for $\theta_f\to0$.
The details of this part of the calculation will be described elsewhere
\cite{dipoleinprep}; here we give only the final result:
\beq
\sigma^{\sub}_{a\gamma\to fX}(p_a,k)
= N_{\mathrm{c},f}\,
\frac{Q_f^2\alpha}{2\pi} \int_0^1\rd x\, {\cal H}(m_f,x,p_a k) \,
\sigma_{a\bar f\to X}(p_a,xk) 
\eeq
with the auxiliary function
\beq
{\cal H}(m_f,x,p_a k) = P_{f\gamma}(x) 
  \ln\biggl(\frac{2p_a k(1-x)^2}{m_f^2}\biggr) +2x(1-x).
\eeq
The contribution $\sigma^{\sub}_{a\gamma\to fX}$ simply has to be added
to the result for the cross section obtained from the integral of the
difference $\sum_{\mathrm{pol}} |\M_{a\gamma\to fX}|^2-|\M_{\sub}|^2$.



\begin{thebibliography}{99} 

\bibitem{Zeller:2001hh}
G.~P.~Zeller {\it et al.}  [NuTeV Collaboration],
Phys.\ Rev.\ Lett.\  {\bf 88} (2002) 091802
[Erratum-ibid.\  {\bf 90} (2003) 239902]
[hep-ex/0110059].

\bibitem{unknown:2004qh}
The LEP Collaborations, the LEP Electroweak Working Group, and SLD 
Electroweak Group and SLD Heavy Flavour Group,
  hep-ex/0412015.

\bibitem{Davidson:2001ji}
S.~Davidson, S.~Forte, P.~Gambino, N.~Rius and A.~Strumia,
JHEP {\bf 0202} (2002) 037
[hep-ph/0112302].

\bibitem{Bernstein:2002sa}
  R.~H.~Bernstein  [NuTeV Collaboration],
  J.\ Phys.\ G {\bf 29} (2003) 1919
  [hep-ex/0210061].

\bibitem{Gambino:2002xp}
P.~Gambino,
hep-ph/0211009.

\bibitem{Giunti:2002nh}
  C.~Giunti and M.~Laveder,
  hep-ph/0202152;\\
%
  W.~Loinaz, N.~Okamura, T.~Takeuchi and L.~C.~R.~Wijewardhana,
  Phys.\ Rev.\ D {\bf 67} (2003) 073012
  [hep-ph/0210193];\\
%
  A.~Kurylov, M.~J.~Ramsey-Musolf and S.~Su,
  Nucl.\ Phys.\ B {\bf 667} (2003) 321
  [hep-ph/0301208];\\
%
  W.~Loinaz, N.~Okamura, S.~Rayyan, T.~Takeuchi and L.~C.~R.~Wijewardhana,
  Phys.\ Rev.\ D {\bf 70} (2004) 113004
  [hep-ph/0403306];\\
%
  O.~Brein, B.~Koch and W.~Hollik,
  hep-ph/0408331;\\
%
  T.~Takeuchi and W.~Loinaz,
  hep-ph/0410201;\\
%
  J.~S.~Ma, J.~M.~Conrad, M.~Sorel and G.~P.~Zeller,
  hep-ex/0501011.

\bibitem{Londergan:2004nr}
  J.~T.~Londergan,
  Nucl.\ Phys.\ Proc.\ Suppl.\  {\bf 141} (2005) 68
  [hep-ph/0408243].

\bibitem{Zeller:2002du}
G.~P.~Zeller {\it et al.}  [NuTeV Collaboration],
Phys.\ Rev.\ D {\bf 65} (2002) 111103
[Erratum-ibid.\ D {\bf 67} (2003) 119902]
[hep-ex/0203004];\\
%
  S.~Kretzer, F.~Olness, J.~Pumplin, D.~Stump, W.~K.~Tung and M.~H.~Reno,
  Phys.\ Rev.\ Lett.\  {\bf 93} (2004) 041802
  [hep-ph/0312322];\\
%
  J.~Alwall and G.~Ingelman,
  Phys.\ Rev.\ D {\bf 70} (2004) 111505
  [hep-ph/0407364];\\
%
  Y.~Ding, R.~G.~Xu and B.~Q.~Ma,
  Phys.\ Lett.\ B {\bf 607} (2005) 101
  [hep-ph/0408292] and
%
  Phys.\ Rev.\ D {\bf 71} (2005) 094014
  [hep-ph/0505153];\\
%
  M.~Wakamatsu,
  Phys.\ Rev.\ D {\bf 71} (2005) 057504
  [hep-ph/0411203];\\
%
  M.~Gl\"uck, P.~Jimenez-Delgado and E.~Reya,
  hep-ph/0501169 and
%
  Phys.\ Rev.\ Lett.\  {\bf 95} (2005) 022002
  [hep-ph/0503103].

\bibitem{McFarland:2003jw}
K.~S.~McFarland and S.~O.~Moch,
hep-ph/0306052.

\bibitem{Martin:2004dh}
A.~D.~Martin, R.~G.~Roberts, W.~J.~Stirling and R.~S.~Thorne,
Eur.\ Phys.\ J.\ C {\bf 39} (2005) 155
[hep-ph/0411040].

\bibitem{Miller:2002xh}
  G.~A.~Miller and A.~W.~Thomas,
  Int.\ J.\ Mod.\ Phys.\ A {\bf 20} (2005) 95
  [hep-ex/0204007];\\
%
  G.~P.~Zeller {\it et al.}  [NuTeV Collaboration],
  hep-ex/0207052;\\
%
  S.~Kovalenko, I.~Schmidt and J.~J.~Yang,
  Phys.\ Lett.\ B {\bf 546} (2002) 68
  [hep-ph/0207158];\\
%
S.~A.~Kulagin,
Phys.\ Rev.\ D {\bf 67} (2003) 091301
[hep-ph/0301045];
%
  hep-ph/0406220 and
%
  Nucl.\ Phys.\ Proc.\ Suppl.\  {\bf 139} (2005) 213
  [hep-ph/0409057];\\
%
  S.~J.~Brodsky, I.~Schmidt and J.~J.~Yang,
  Phys.\ Rev.\ D {\bf 70} (2004) 116003
  [hep-ph/0409279];\\
%
  M.~Hirai, S.~Kumano and T.~H.~Nagai,
  Phys.\ Rev.\ D {\bf 71} (2005) 113007
  [hep-ph/0412284];\\
%
  S.~A.~Kulagin and R.~Petti,
  hep-ph/0412425.

\bibitem{Kretzer:2003iu}
S.~Kretzer and M.~H.~Reno,
hep-ph/0307023;\\
%
B.~A.~Dobrescu and R.~K.~Ellis,
hep-ph/0310154.

\bibitem{Marciano:pb}
W.~J.~Marciano and A.~Sirlin,
Phys.\ Rev.\ D {\bf 22} (1980) 2695
[Erratum-ibid.\ D {\bf 31} (1985) 213] and
Nucl.\ Phys.\ B {\bf 189} (1981) 442.

\bibitem{Bardin:1986bc}
D.~Y.~Bardin and V.~A.~Dokuchaeva,
{\it On The Radiative Corrections To The Neutrino Deep Inelastic Scattering},
JINR-E2-86-260.

\bibitem{Diener:2003ss}
K.~P.~O.~Diener, S.~Dittmaier and W.~Hollik,
Phys.\ Rev.\ D {\bf 69} (2004) 073005
[hep-ph/0310364].

\bibitem{Arbuzov:2004zr}
A.~B.~Arbuzov, D.~Y.~Bardin and L.~V.~Kalinovskaya,
hep-ph/0407203.

\bibitem{Petti:2004wy}
  R.~Petti  [NOMAD Collaboration],
  hep-ex/0411032.

\bibitem{LlewellynSmith:ie}
C.~H.~Llewellyn Smith,
Nucl.\ Phys.\ B {\bf 228} (1983) 205.

\bibitem{Paschos:1972kj}
E.~A.~Paschos and L.~Wolfenstein,
Phys.\ Rev.\ D {\bf 7} (1973) 91.

\bibitem{Denner:1993kt}
A.~Denner,
Fortsch.\ Phys.\  {\bf 41} (1993) 307.

\bibitem{Bohm:rj}
M.~B\"ohm, H.~Spiesberger and W.~Hollik,
Fortsch.\ Phys.\  {\bf 34} (1986) 687;\\
W.~F.~Hollik,
Fortsch.\ Phys.\  {\bf 38} (1990) 165.

\bibitem{Sirlin:1980nh}
A.~Sirlin,
Phys.\ Rev.\ D {\bf 22} (1980) 971.

\bibitem{Consoli:1989pc}
M.~Consoli, W.~Hollik and F.~Jegerlehner,
CERN-TH-5527-89,
{\it Presented at Workshop on Z Physics at LEP} and
%
Phys.\ Lett.\ B {\bf 227} (1989) 167.

\bibitem{Fleischer:1993ub}
J.~Fleischer, O.~V.~Tarasov and F.~Jegerlehner,
Phys.\ Lett.\ B {\bf 319} (1993) 249.

\bibitem{KLN}
T.~Kinoshita,
J.\ Math.\ Phys.\ {\bf 3} (1962) 650; \\
T.~D.~Lee and M.~Nauenberg,
Phys.\ Rev.\ {\bf 133} (1964) B1549.

\bibitem{sf}
E.~A.~Kuraev and V.~S.~Fadin,
Sov.\ J.\ Nucl.\ Phys.\  {\bf 41} (1985) 466
[Yad.\ Fiz.\  {\bf 41} (1985) 733];\\
%
G.~Altarelli and G.~Martinelli, in {\it Physics at LEP},
eds. J.~Ellis and R.~Peccei, (CERN 86-02, Geneva, 1986), Vol.~1, p.~47;\\
%
O.~Nicrosini and L.~Trentadue,
Phys.\ Lett.\ B {\bf 196} (1987) 551 and
%
Z.\ Phys.\ C {\bf 39} (1988) 479;\\
%
F.~A.~Berends, W.~L.~van Neerven and G.~J.~H.~Burgers,
Nucl.\ Phys.\ B {\bf 297} (1988) 429
[Erratum-ibid.\ B {\bf 304} (1988) 921];\\
%
A.~B.~Arbuzov,
Phys.\ Lett.\ B {\bf 470} (1999) 252
[hep-ph/9908361].

\bibitem{Humpert:1980uv}
  B.~Humpert and W.~L.~van Neerven,
  Nucl.\ Phys.\ B {\bf 184} (1981) 225.

\bibitem{Kripfganz:1988bd}
J.~Kripfganz and H.~Perlt,
Z.\ Phys.\ C {\bf 41} (1988) 319;\\
H.~Spiesberger,
Phys.\ Rev.\ D {\bf 52} (1995) 4936
[hep-ph/9412286];\\
%
  M.~Roth and S.~Weinzierl,
  Phys.\ Lett.\ B {\bf 590} (2004) 190
  [hep-ph/0403200].

\bibitem{privcom}
V.~Lendermann, H.~Spiesberger, J.~Stirling and A.~Tapper, 
private communications.

\bibitem{Eidelman:2004wy}
S.~Eidelman {\it et al.}  [Particle Data Group Collaboration],
Phys.\ Lett.\ B {\bf 592} (2004) 1.

\bibitem{Azzi:2004rc}
P.~Azzi {\it et al.}  [CDF and D0 Collaborations, 
and Tevatron Electroweak Working Group],
hep-ex/0404010.

\bibitem{Jegerlehner:2001ca}
F.~Jegerlehner,
DESY 01-029, LC-TH-2001-035 [hep-ph/0105283].

\bibitem{Catani:1996jh}
S.~Catani and M.~H.~Seymour,
Phys.\ Lett.\ B {\bf 378} (1996) 287
[hep-ph/9602277] and
Nucl.\ Phys.\ B {\bf 485} (1997) 291
[Erratum-ibid.\ B {\bf 510} (1997) 291]
[hep-ph/9605323].

\bibitem{Dittmaier:2000mb}
S.~Dittmaier,
Nucl.\ Phys.\ B {\bf 565} (2000) 69
[hep-ph/9904440].

\bibitem{Roth:1999kk}
M.~Roth,
PhD thesis, ETH Z\"urich No. 13363 (1999),
hep-ph/0008033.

\bibitem{Bredenstein:2005zk}
  A.~Bredenstein, S.~Dittmaier and M.~Roth,
  hep-ph/0506005, to appear in Eur.\ Phys.\ J.\ C.

\bibitem{dipoleinprep}
S.~Dittmaier, in preparation.

\bibitem{Faisst:2003px}
  M.~Faisst, J.~H.~Kuhn, T.~Seidensticker and O.~Veretin,
  Nucl.\ Phys.\ B {\bf 665} (2003) 649
  [hep-ph/0302275].

\bibitem{Dittmaier:2001ay}
S.~Dittmaier and M.~Kr\"amer,
Phys.\ Rev.\ D {\bf 65} (2002) 073007
[hep-ph/0109062].

\bibitem{Catani:2002hc}
  S.~Catani, S.~Dittmaier, M.~H.~Seymour and Z.~Trocsanyi,
  Nucl.\ Phys.\ B {\bf 627} (2002) 189
  [hep-ph/0201036].

\end{thebibliography}
\end{document}